\documentclass[useAMS,usenatbib]{mn2e}
\usepackage[dvips]{graphicx}
\usepackage[english]{babel}
\usepackage{amsmath}
\usepackage{amssymb}
\usepackage{textcomp}
\usepackage{natbib}

\title[Gas properties in and around haloes]{Properties of gas in and around galaxy haloes}
\author[F. van de Voort and J. Schaye]{Freeke van de Voort$^{1}$\thanks{E-mail:fvdvoort@strw.leidenuniv.nl} and Joop~Schaye$^1$\\
$^{1}$Leiden Observatory, Leiden University, P.O. Box 9513, 2300 RA, Leiden, The Netherlands}

\begin{document}

\date{Accepted 2012 March 15. Received 2012 February 19; in original form 2011 November 21}

\pagerange{\pageref{firstpage}--\pageref{lastpage}} \pubyear{2011}

\maketitle

\label{firstpage}

\begin{abstract}

We study the properties of gas inside and around galaxy haloes as a function of radius and halo mass, focussing mostly on $z=2$, but also showing some results for $z=0$. For this purpose, we use a suite of large cosmological, hydrodynamical simulations from the OverWhelmingly Large Simulations project. The properties of cold- and hot-mode gas, which we separate depending on whether the temperature has been higher than $10^{5.5}$~K while it was extragalactic, are clearly distinguishable in the outer parts of massive haloes (virial temperatures $\gg 10^5$~K). The differences between cold- and hot-mode gas resemble those between inflowing and outflowing gas. The cold-mode gas is mostly confined to clumpy filaments that are approximately in pressure equilibrium with the diffuse, hot-mode gas. Besides being colder and denser, cold-mode gas typically has a much lower metallicity and is much more likely to be infalling. However, the spread in the properties of the gas is large, even for a given mode and a fixed radius and halo mass, which makes it impossible to make strong statements about individual gas clouds. Metal-line cooling causes a strong cooling flow near the central galaxy, which makes it hard to distinguish gas accreted through the cold and hot modes in the inner halo. Stronger feedback results in larger outflow velocities and pushes hot-mode gas to larger radii. The gas properties evolve as expected from virial arguments, which can also account for the dependence of many gas properties on halo mass. We argue that cold streams penetrating hot haloes are observable as high-column density H\,\textsc{i} Lyman-$\alpha$ absorption systems in sightlines near massive foreground galaxies. 
\end{abstract}


\section{Introduction}

The gaseous haloes around galaxies grow by accreting gas from their surroundings, the intergalactic medium, which is the main reservoir for baryons. The galaxies themselves grow by accreting gas from their haloes, from which they can form stars. Some of the gas is, however, returned to the circumgalactic medium by galactic winds driven by supernovae (SNe) or active galactic nuclei (AGN) and by dynamical processes such as tidal or ram pressure forces. Such interactions between the different gas phases are essential for galaxy formation and evolution.

The physical state of the gas in and around haloes will determine how fast galaxies grow. Quantifying and understanding the properties of the gas is therefore vital for theories of galaxy formation.
It is also crucial for making predictions and for the interpretation of observations as the physical state of the gas determines how much light is absorbed and emitted. 

Theoretical and computational studies of the accretion of gas onto galaxies have revealed the existence of two distinct modes. In the first mode the inflowing gas experiences an accretion shock as it collides with the hot, hydrostatic halo near the virial radius. At that point it is shock-heated to temperatures similar to the virial value and typically remains part of the hot halo for longer than a dynamical time. If it reaches a sufficiently high density, it can cool radiatively and settle into a disc \citep[e.g.][]{Rees1977, White1978,Fall1980}. This mode is referred to as `hot-mode accretion' \citep{Katz2003, Keres2005}. If, on the other hand, the cooling time of the gas is short compared to the dynamical time, which is the case for haloes of sufficiently low mass, a hot halo is unable to form and the accreting gas will not go through a shock near the virial radius. The accretion rate then depends on the infall rate instead of on the cooling rate \citep{White1991, Birnboim2003, Dekel2006}. Additionally, simulations have shown that much of the gas enters the halo along dense filaments or in clumps, which gives rise to short cooling times, even in the presence of a hot, hydrostatic halo. This denser gas does not go through an accretion shock near the virial radius and will therefore remain cold until it accretes onto the central galaxy or is hit by an outflow \citep[e.g.][]{Keres2005, Dekel2009a,Voort2011a}. We refer to this mode as `cold-mode accretion' \citep{Katz2003, Keres2005}.

Hot- and cold-mode accretion play very different roles in the formation of galaxies and their gaseous haloes \citep{Keres2005, Ocvirk2008, Keres2009a, Keres2009b, Brooks2009, Dekel2009a, Crain2010a, Voort2011a, Voort2011b, Powell2011, Faucher2011}. It has been shown that cold-mode accretion is more important at high redshift, when the density of the Universe is higher. Hot-mode accretion dominates the fuelling of the gaseous haloes of high-mass systems \citep[halo mass $> 10^{12} \, {\rm M_\odot}$; e.g.\ ][]{Ocvirk2008, Voort2011a}. The importance of hot-mode accretion is much reduced when considering accretion onto galaxies (as opposed to haloes) \citep{Keres2009a, Voort2011a}. At $z\ge1$ all galaxies accrete more than half of their material in the cold mode, although the contribution of hot-mode accretion is not negligible for high-mass haloes. Cold-mode accretion provides most of the fuel for star formation and shapes the cosmic star formation rate
density \citep{Voort2011b}. 

\citet{Voort2011a,Voort2011b} investigated the roles of feedback mechanisms on the gas accretion. They found that while the inclusion of metal-line cooling has no effect on the accretion onto haloes, it does increase the accretion rate onto galaxies, because it decreases the cooling time of the hot halo gas. Feedback from SNe and AGN can reduce the accretion rates onto haloes by factors of a few, but accretion onto galaxies is suppressed by up to an order of magnitude. The inclusion of AGN feedback is particularly important for suppressing hot-mode accretion onto galaxies, because it is mainly effective in high-mass haloes and because diffuse gas is more susceptible to outflows. 

Hot, hydrostatic halo gas is routinely studied using X-ray observations of galaxy groups and clusters and has perhaps even been detected around individual galaxies \citep[e.g.][]{Crain2010a, Crain2010b, Anderson2011}. As of yet, there is no direct observational evidence for cold-mode accretion, even though there are claims of individual detections in H\,\textsc{i} absorption based on the low metallicity and proximity to a galaxy of the absorption system \citep{Ribaudo2011, Giavalisco2011}. Cosmological simulations can reproduce the observed H\,\textsc{i} column density distribution \citep{Altay2011}. They show that cold-mode accretion is responsible for much of the observed high column density H\,\textsc{i} absorption at $z\sim 3$. In particular, most of the detected Lyman limit and low column density damped Lyman-$\alpha$ absorption may arise in cold accretion streams \citep{Fumagalli2011a, Voort2011c}.

It has also been claimed that the diffuse Lyman-$\alpha$ emission detected around some high-redshift galaxies is powered by cold accretion \cite[e.g.][]{Fardal2001, Dijkstra2009, Goerdt2010, Rosdahl2012}, but both simulations and observations indicate that the emission is more likely scattered light from central H\,\textsc{ii} regions \cite[e.g.][]{Furlanetto2005, Faucher2010, Steidel2010, Hayes2011, Rauch2011}.

The temperature is, however, not the only difference between the two accretion modes. In this paper we use the suite of cosmological hydrodynamical simulations from the OverWhelmingly Large Simulations project \citep[OWLS; ][]{Schaye2010} to investigate other physical properties, such as the gas density, pressure, entropy, metallicity, radial peculiar velocity, and accretion rate of the gas in the two modes. We will study the dependence of gas properties on radius for haloes of total mass $\sim 10^{12}~{\rm M}_\odot$ and the dependence on halo mass of the properties of gas just inside the virial radius.
Besides contrasting the hot and cold accretion modes, we will also distinguish between inflowing and outflowing gas. While most of our results will be presented for $z=2$, when both hot- and cold-mode accretion are important for haloes of mass $\sim 10^{12}~{\rm M}_\odot$, we will also present some results for $z=0$, which are therefore directly relevant for observations of gas around the Milky Way. We will make use of the different OWLS runs to investigate how the results vary with the efficiency of the feedback and the cooling. 

This paper is organized as follows. The simulations are described in Section~\ref{sec:sim}, including the model variations, the way in which haloes are identified, and our method for distinguishing gas accreting in the hot and cold modes. In Sections~\ref{sec:properties} and \ref{sec:mass} we study the radial profiles and the dependence on halo mass, respectively. In Section~\ref{sec:inout} we discuss the difference in physical properties between inflowing and outflowing gas. We assess the effect of metal-line cooling and feedback from SNe and AGN on the gas properties in Section~\ref{sec:SNAGN}. In Section~\ref{sec:z0} we study the properties of gas around Milky Way-sized galaxies at $z=0$. Finally, we discuss and summarize our conclusions in Section~\ref{sec:concl}.

\section{Simulations} \label{sec:sim}

To investigate the gas properties in and around haloes, we make use of simulations taken from the OWLS project \citep{Schaye2010}, which consists of a large number of cosmological simulations, with varying (subgrid) physics. Here, we make use of a subset of these simulations. We first summarize the reference simulation, from which we derive our main results. The other simulations are described in Section~\ref{sec:var}. For a full description of the simulations, we refer the reader to \citet{Schaye2010}. Here, we will only summarize their main properties.

We use a modified version of \textsc{gadget-3} \citep[last described in][]{Springel2005}, a smoothed particle hydrodynamics (SPH) code that uses the entropy formulation of SPH \citep{Springel2002}, which conserves both energy and entropy where appropriate. 

All the cosmological simulations used in this work assume a $\Lambda$CDM cosmology with parameters derived from the WMAP year~3 data, $\Omega_\mathrm{m} = 1 - \Omega_\Lambda = 0.238$, $\Omega_\mathrm{b} = 0.0418$, $h = 0.73$, $\sigma_8 = 0.74$, $n = 0.951$ \citep{Spergel2007}. These values are consistent\footnote{The most significant discrepancy is in $\sigma_8$, which is 8 per cent, or $2.3\sigma$, lower than the value favoured by the WMAP 7-year data.} with the WMAP year~7 data \citep{Komatsu2011}. 
The primordial abundances are $X = 0.752$ and $Y = 0.248$, where $X$ and $Y$ are the mass fractions of hydrogen and helium, respectively. 

\begin{table*}
\caption{\label{tab:res} \small Simulation parameters: simulation identifier, comoving box size ($L_\mathrm{box}$), number of dark matter particles ($N$, the number of baryonic particles is equal to the number of dark matter particles), mass of dark matter particles ($m_\mathrm{DM}$), initial mass of gas particles ($m_\mathrm{gas}^\mathrm{initial}$), number of haloes with $10^{11.5}$~M$_\odot<M_\mathrm{halo}<10^{12.5}$~M$_\odot$, and number of haloes with more than 100~dark matter particles.}
\begin{tabular}[t]{lrrrrrr}
\hline
\hline \\[-3mm]
simulation & $L_\mathrm{box}$ & $N$ & $m_\mathrm{DM}$ & $m_\mathrm{gas}^\mathrm{initial}$ & number of haloes with & number of resolved haloes \\
 & ($h^{-1}$Mpc) & & (M$_\odot$) & (M$_\odot$) & $10^{11.5}$~M$_\odot<M_\mathrm{halo}<10^{12.5}$~M$_\odot$ & at $z=2$ \\
\hline \\[-4mm]
\emph{L100N512} & 100 & 512$^3$ & $5.56\times 10^8$ & $1.19 \times 10^8$ & 4407 ($z=2$) & 32167 \\
\emph{L050N512} & {50} & 512$^3$ & $6.95\times 10^7$ & $1.48\times 10^7$ & 518 ($z=2$); 1033 ($z=0$) & 32663\\
\emph{L025N512} & {25} & 512$^3$ & $8.68\times 10^6$ & $1.85\times 10^6$ & 59 ($z=2$)& 25813 \\
\hline
\end{tabular}
\end{table*}

A cubic volume with periodic boundary conditions is defined, within which the mass is distributed over $N^3$ dark matter and as many gas particles. The box size (i.e.\ the length of a side of the simulation volume) of the simulations used in this work are 25, 50, and 100~$h^{-1}$Mpc, with $N=512$. The (initial) particle masses for baryons and dark matter are $1.5\times10^7(\frac{L_\mathrm{box}}{50\ h^{-1}\mathrm{Mpc}})^3$~M$_\odot$ and $7.0\times10^7(\frac{L_\mathrm{box}}{50\ h^{-1}\mathrm{Mpc}})^3$~M$_\odot$, respectively, and are listed in Table~\ref{tab:res}. We use the notation \emph{L***N\#\#\#}, where \emph{***} indicates the box size in comoving Mpc$/h$ and \emph{\#\#\#} the number of particles per dimension.
The gravitational softening length is initially 3.9~$(\frac{L_\mathrm{box}}{50\ h^{-1}\mathrm{Mpc}})$~$h^{-1}$ comoving kpc, i.e.\ 1/25 of the mean dark matter particle separation, but we imposed a maximum of 1~$(\frac{L_\mathrm{box}}{50\ h^{-1}\mathrm{Mpc}})$~$h^{-1}$kpc proper. We use simulation \emph{REF\_L050N512} for our main results. The \emph{L025N512} simulations are used for images, for comparisons between simulations with different subgrid physics, and for resolution tests. The \emph{L100N512} run is only used for the convergence tests shown in the appendix.

The abundances of eleven elements (hydrogen, helium, carbon, nitrogen, oxygen, neon, magnesium, silicon, sulphur, calcium, and iron) released by massive stars (type II SNe and stellar winds) and intermediate mass stars (type Ia SNe and asymptotic giant branch stars) are followed as described in \citet{Wiersma2009b}.
We assume the stellar initial mass function (IMF) of \citet{Chabrier2003}, ranging from 0.1 to 100~M$_\odot$. As described in \citet{Wiersma2009a}, radiative cooling and heating are computed element-by-element in the presence of the cosmic microwave background radiation and the \citet{Haardt2001} model for the UV/X-ray background from galaxies and quasars. The gas is assumed to be optically thin and in (photo)ionization equilibrium.

Star formation is modelled according to the recipe of \citet{Schaye2008}. The Jeans mass cannot be resolved in the cold, interstellar medium (ISM), which could lead to artificial fragmentation \citep[e.g.][]{Bate1997}. Therefore, a polytropic equation of state $P_\mathrm{tot}\propto\rho_\mathrm{gas}^{4/3}$ is implemented for densities exceeding $n_\mathrm{H}=0.1$~cm$^{-3}$, where $P_\mathrm{tot}$ is the total pressure and $\rho_\mathrm{gas}$ the density of the gas. This equation of state makes the Jeans mass, as well as the ratio of the Jeans length and the SPH smoothing kernel, independent of the density. Gas particles whose proper density exceeds $n_\mathrm{H}\ge0.1$~cm$^{-3}$ while they have temperatures $T\le10^5$~K are moved on to this equation of state and can be converted into star particles. The star formation rate per unit mass depends on the gas pressure and is set to reproduce the observed Kennicutt-Schmidt law \citep{Kennicutt1998}.

Feedback from star formation is implemented using the prescription of \citet{Vecchia2008}. About 40 per cent of the energy released by type II SNe is injected locally in kinetic form. The rest of the energy is assumed to be lost radiatively. Each gas particle within the SPH smoothing kernel of the newly formed star particle has a probability of being kicked. For the reference model, the mass loading parameter $\eta = 2$, meaning that, on average, the total mass of the particles being kicked is twice the mass of the star particle formed. Because the winds sweep up surrounding material, the effective mass loading can be much higher. The initial wind velocity is 600~km\,s$^{-1}$ for the reference model. \citet{Schaye2010} showed that these parameter values yield a peak global star formation rate density that agrees with observations.

\subsection{Variations} \label{sec:var}

\begin{table}
\caption{\label{tab:owls} \small Simulation parameters: simulation identifier, cooling including metals (Z cool), initial wind velocity ($v_\mathrm{wind}$), initial wind mass loading ($\eta$), AGN feedback included (AGN). Differences from the reference model are indicated in bold face.}
\begin{tabular}[t]{lcrcr}
\hline
\hline \\[-3mm]
simulation & Z cool & $v_\mathrm{wind}$ & $\eta$  & AGN \\
  & & (km\,s$^{-1}$) & & \\
\hline \\[-4mm]
\emph{REF} & yes & 600 & 2 & no \\
\emph{NOSN\_NOZCOOL} & \textbf{no} & \textbf{\ \ \ 0} & \textbf{0} & no \\
\emph{NOZCOOL} & \textbf{no} & 600 & 2 & no \\
\emph{WDENS} & yes &  \multicolumn{2}{r}{\textbf{density dependent}} & no \\
\emph{AGN} & yes & 600 & 2 & \textbf{yes} \\
\hline
\end{tabular}
\end{table}

To investigate the effect of feedback and metal-line cooling, we have performed a suite of simulations in which the subgrid prescriptions are varied. These are listed in Table~\ref{tab:owls}.

The importance of metal-line cooling can be demonstated by comparing the reference simulation (\emph{REF}) to a simulation in which primordial abundances are assumed when calculating the cooling rates (\emph{NOZCOOL}). We also performed a simulation in which both cooling by metals and feedback from SNe were omitted (\emph{NOSN\_NOZCOOL}). To study the effect of SN feedback, this simulation can be compared to (\emph{NOZCOOL}).

In massive haloes the pressure of the ISM is too high for winds with velocities of 600~km\,s$^{-1}$ to blow the gas out of the galaxy \citep{Vecchia2008}. To make the winds effective at higher halo masses, the velocity can be scaled with the local sound speed, while adjusting the mass loading so as to keep the energy injected per unit stellar mass constant at $\approx 40$ per cent (\emph{WDENS}). 

Finally, we have included AGN feedback (\emph{AGN}). Black holes grow via mergers and gas accretion and inject 1.5 per cent of the rest-mass energy of the accreted gas into the surrounding matter in the form of heat. The model is based on the one introduced by \citet{Springeletal2005} and is described and tested in \citet{Booth2009}, who also demonstrate that the simulation reproduces the observed mass density in black holes and the observed scaling relations between black hole mass and central stellar velocity dispersion and between black hole mass and stellar mass. \citet{McCarthy2010} have shown that model \emph{AGN} reproduces the observed stellar mass fractions, star formation rates, and stellar age distributions in galaxy groups, as well as the thermodynamic properties of the intragroup medium.

\subsection{Identifying haloes} \label{sec:halo}

The first step towards finding gravitationally bound structures is to identify dark matter haloes. These can be found using a Friends-of-Friends (FoF) algorithm. If the separation between two dark matter particles is less than 20 per cent of the average separation (the linking length $b=0.2$), they are placed in the same group. Baryonic particles are linked to a FoF halo if their nearest dark matter neighbour is in that halo. We then use \textsc{subfind} \citep{Dolag2009} to find the most bound particle of a FoF halo, which serves as the halo centre. In this work we use a spherical overdensity criterion, considering all the particles in the simulation. We compute the virial radius, $R_\mathrm{vir}$, within which the average density agrees with the prediction of the top-hat spherical collapse model in a $\Lambda$CDM cosmology \citep{Bryan1998}. At $z=2$ this corresponds to a density of $\rho = 169\langle\rho\rangle$.

We include only haloes containing more than 100 dark matter particles in our analysis, corresponding to a minimum dark matter halo mass of $M_\mathrm{halo}=10^{10.7}$, $10^{9.8}$, and $10^{8.9}$~M$_\odot$ in the 100, 50, and 25~$h^{-1}$Mpc box, respectively.
For these limits our mass functions agree very well with the \citet{Sheth1999} fit. Table~\ref{tab:res} lists, for each simulation of the reference model, the number of haloes with mass $10^{11.5}$~M$_\odot<M_\mathrm{halo}<10^{12.5}$~M$_\odot$ and the number of haloes with more than 100 dark matter particles.

\subsection{Hot- and cold-mode gas}

During the simulations the maximum past temperature, $T_\mathrm{max}$, was stored in a separate variable. The variable was updated for each SPH particle at every time step for which the temperature was higher than the previous maximum past temperature. The artificial temperature the particles obtain when they are on the equation of state (i.e.\ when they are part of the unresolved multiphase ISM) was ignored in this process. This may cause us to underestimate the maximum past temperature of gas that experienced an accretion shock at densities $n_{\rm H} > 0.1 ~{\rm cm}^{-3}$. Ignoring such shocks is, however, consistent with our aims, as we are interested in the maximum temperature reached \emph{before} the gas entered the galaxy. Note, however, that the maximum past temperature of some particles may reflect shocks in outflowing rather than accreting gas.

Another reason why $T_\mathrm{max}$ may underestimate the true maximum past temperature, is that in SPH simulations a shock is smeared out over a few smoothing lengths, leading to in-shock cooling \citep{Hutchings2000}. If the cooling time is of the order of, or smaller than, the time step, then the maximum temperature will be underestimated. \citet{Creasey2011} have shown that a particle mass of $10^6$~M$_\odot$ is sufficient to avoid numerical overcooling of accretion shocks onto haloes, like in our high-resolution simulations (\emph{L025N512}). The Appendix shows that our lower-resolution simulations give very similar results.

Even with infinite resolution, the post-shock temperatures may, however, not be well defined. Because electrons and protons have different masses, they will have different temperatures in the post-shock gas and it may take some time before they equilibrate through collisions or plasma effects. We have ignored this complication. Another effect, which was also not included in our simulation, is that shocks may be preceded by the radiation from the shock, which may affect the temperature evolution. Disregarding these issues, \citet{Voort2011a} showed that the distribution of $T_\mathrm{max}$ is bimodal and that a cut at $T_\mathrm{max}=10^{5.5}$~K naturally divides the gas into cold- and hot-mode accretion and that it produces similar results as studies based on adaptive mesh refinement simulations \citep{Ocvirk2008}. This $T_\mathrm{max}$ threshold was chosen because the cooling function peaks at $10^{5-5.5}$~K \citep[e.g.][]{Wiersma2009a}, which results in a minimum in the temperature distribution. Additionally, the UV background can only heat gas to about $10^5$~K, which is therefore characteristic for cold-mode accretion. In this work we use the same $T_\mathrm{max}=10^{5.5}$~K threshold to separate the cold and hot modes.

\section{Physical properties: dependence on radius} \label{sec:properties}

\begin{figure*}
\center
\includegraphics[scale=0.7]{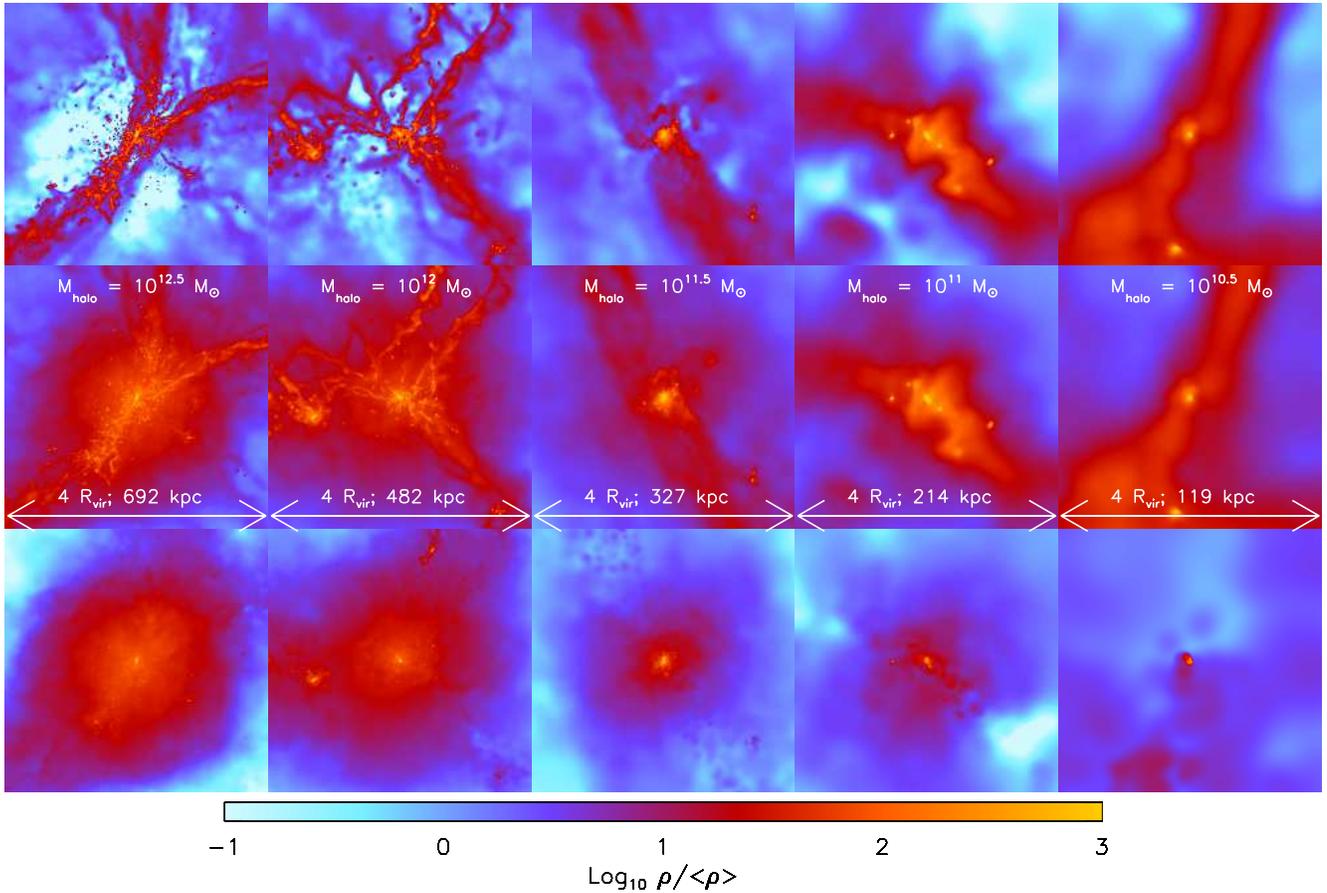}
\caption {\label{fig:dens} Gas overdensity at $z=2$ in and around haloes with, from left to right, $M_\mathrm{halo}=10^{12.5}$, $10^{12}$, $10^{11.5}$, $10^{11}$, and $10^{10.5}$~M$_\odot$ taken from the simulation \emph{REF\_L025N512}. All images show projections of the gas density in cubes of 4 virial radii on a side. The proper sizes of the images are indicated in the panels of the middle row. In the middle column all gas was included. In the top (bottom) row we have only included cold- (hot-)mode gas, i.e.\ gas with $T_\mathrm{max}<10^{5.5}$~K ($T_\mathrm{max}\ge10^{5.5}$~K). The filaments, streams, and dense clumps consist of gas that has never been heated to temperatures greater than $10^{5.5}$~K.}
\end{figure*}

The gas in the Universe is distributed in a cosmic web of sheets, filaments, and haloes. The filaments also affect the structure of the haloes that reside inside them or at their intersections. At high redshift, cold, narrow streams penetrate hot haloes and feed galaxies efficiently \citep[e.g.][]{Keres2005, Dekel2006, Agertz2009, Ceverino2010, Voort2011a}. The middle row of Fig.~\ref{fig:dens} shows the overdensity in several haloes with different masses, ranging from $M_\mathrm{halo}=10^{12.5}$ (left panel) to $10^{10.5}$~M$_\odot$ (right panel), taken from the high-resolution reference simulation (\emph{REF\_L025N512}) at $z=2$.
Each image is four virial radii on a side, so the physical scale decreases with decreasing halo mass, as indicated in the middle panels. To illustrate the morphologies of gas that was accreted in the different modes, we show the density of the cold- and hot-mode gas separately in the top and bottom panels, respectively. The spatial distribution is clearly different. Whereas the cold-mode gas shows clear filaments and many clumps, the hot-mode gas is much more spherically symmetric and smooth, particularly for the higher halo masses. The filaments become broader, relative to the size of the halo, for lower mass haloes. In high-mass haloes, the streams look disturbed and some fragment into small, dense clumps, whereas they are broad and smooth in low-mass haloes. Cold-mode accretion is clearly possible in haloes that are massive enough to have well-developed virial shocks if the density of the accreting gas is high, which is the case when the gas accretes along filaments or in clumps.

\begin{figure*}
\center
\includegraphics[scale=0.83]{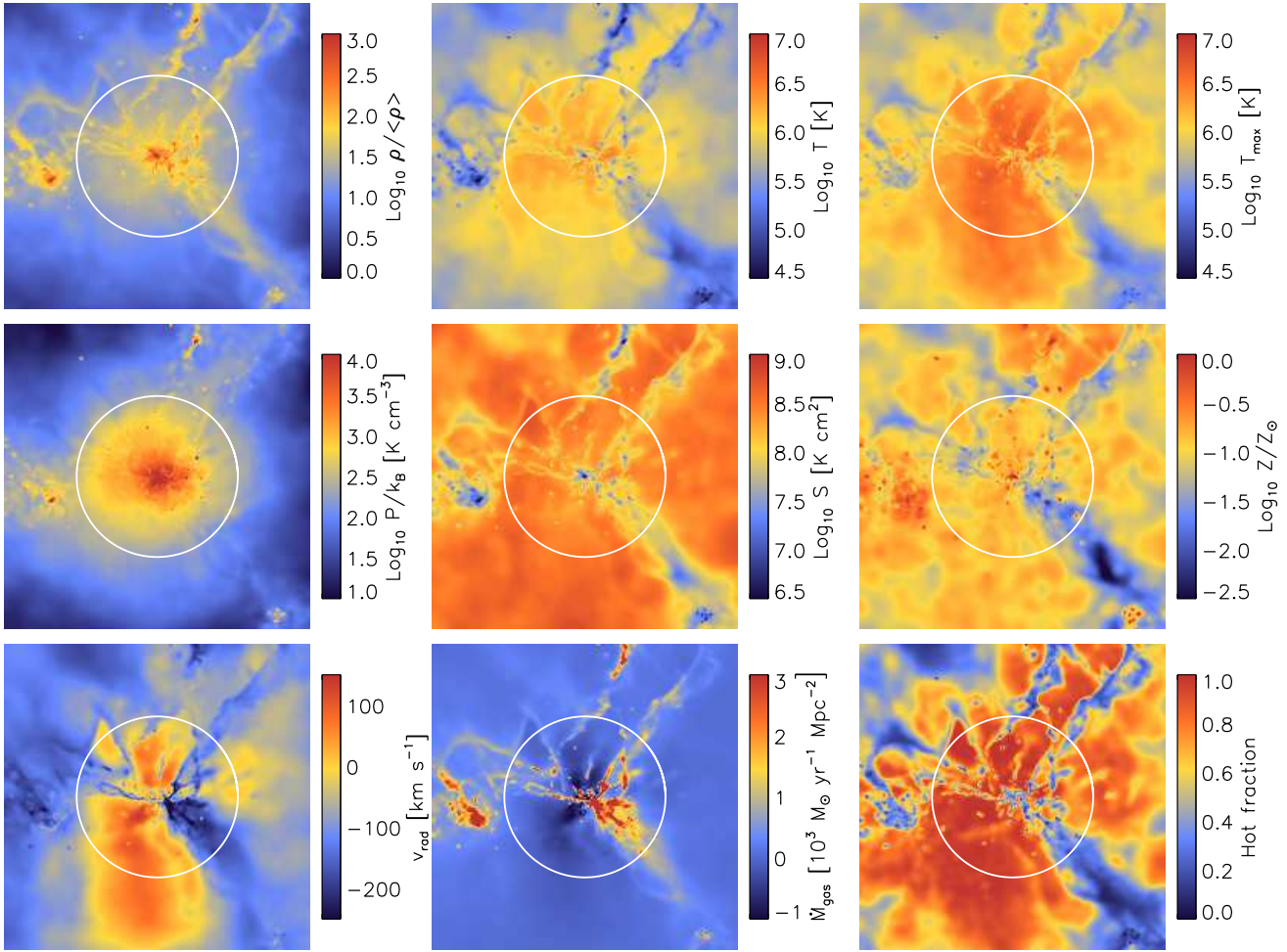}
\caption {\label{fig:halo} From the top-left to the bottom-right: gas overdensity, temperature, maximum past temperature, pressure, entropy, metallicity, radial peculiar velocity, radial mass flux (in solar mass per year per proper Mpc$^2$), and hot fraction in a cubic 1~$h^{-1}$~comoving Mpc region centred on a halo of $M_\mathrm{halo}\approx10^{12}$~M$_\odot$ at $z=2$ taken from the \emph{REF\_L025N512} simulation. The white circles indicate the virial radius.}
\end{figure*}

Fig.~\ref{fig:halo} shows several physical quantities for the gas in a cubic 1~$h^{-1}$ comoving Mpc region, which is about four times the virial radius, centred on the $10^{12}$~M$_\odot$ halo from Fig.~\ref{fig:dens}. These properties are (from the top-left to the bottom-right): gas overdensity, temperature, maximum past temperature, pressure, entropy, metallicity, radial peculiar velocity, radial mass flux, and finally the ``hot fraction'' which we define as the mass fraction of the gas that was accreted in the hot mode (i.e.\ that has $T_\mathrm{max}\ge10^{5.5}$~K). The properties are mass-weighted and projected along the line of sight. The virial radius is 264~$h^{-1}$ comoving kpc and is indicated by the white circles.

\begin{figure*}
\center
\includegraphics[scale=0.65]{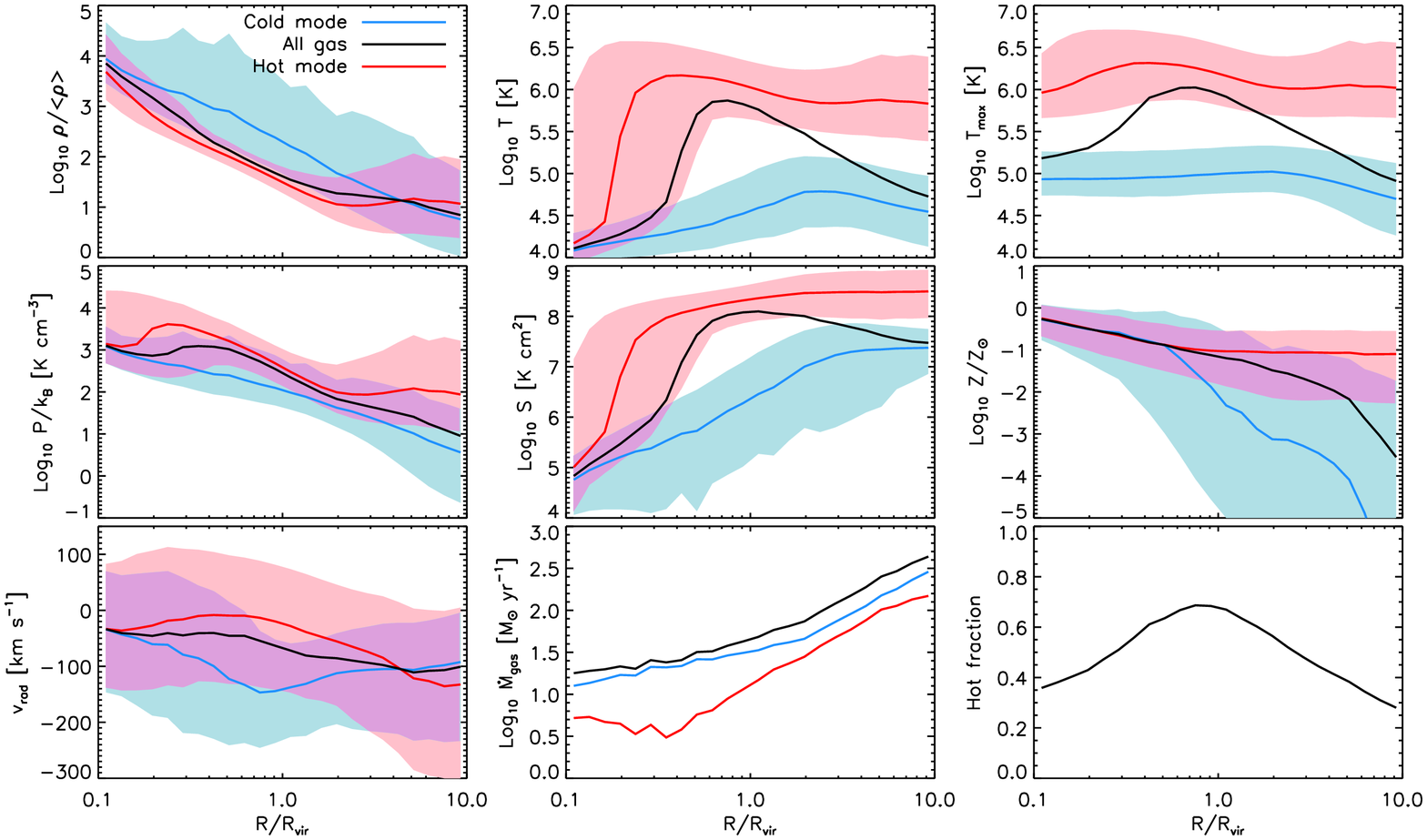}
\caption {\label{fig:haloradz2} Properties of gas in haloes with $10^{11.5}$~M$_\odot<M_\mathrm{halo}<10^{12.5}$~M$_\odot$ at $z=2$ as a function of radius for all (black curves), hot-mode (red curves), and cold-mode (blue curves) gas. Results are shown for the simulation \emph{REF\_L050N512}. Shaded regions show values within the 16th and 84th percentiles, i.e., the $\pm 1\sigma$ scatter around the median. From the top-left to the bottom-right, the different panels show the mass-weighted median gas overdensity, temperature, maximum past temperature, pressure, entropy, metallicity, radial peculiar velocity, the mean accretion rate, and the mean mass fraction of hot-mode gas, respectively. 
}
\end{figure*}

In Fig.~\ref{fig:haloradz2} we show the same quantities as in Fig.~\ref{fig:halo} as a function of radius for the haloes with $10^{11.5}$~M$_\odot<M_\mathrm{halo}<10^{12.5}$~M$_\odot$ at $z=2$ in simulation \emph{REF\_L050N512}. The black curves show the median values for all gas, except for the last two panels which show the mean values. The red (blue) curves show the median or mean values for hot- (cold-)mode gas, i.e.\ gas with maximum past temperatures above (below) $10^{5.5}$~K. The shaded regions show values within the 16th and 84th percentiles. Hot-mode gas at radii larger than $2R_\mathrm{vir}$ is dominated by gas associated with other haloes and/or large-scale filaments.

All the results we show are weighted by mass. In other words, we stacked all 518 haloes in the selected mass range using $R/R_{\rm vir}$ as the radial coordinate. The black curves in Fig.~\ref{fig:haloradz2} (except for the last two panels) then show the values of the corresponding property (e.g.\ the gas overdensity in the top-left panel) that divide the total mass in each radial bin in half, i.e.\ half the mass lies above the curve. We have done the same analysis for volume-weighted quantities by computing, as a function of radius, the values of each property that divides the total volume, i.e.\ the sum of $m_\mathrm{gas}/\rho$, in half but we do not show the results. The volume is completely dominated by hot-mode gas out to twice the virial radius, reaching 50 per cent at $3R_\mathrm{vir}$. Even though the volume-weighted hot fraction is very different, the properties of the gas and the differences between the properties of hot- and cold-mode gas are similar if we weigh by volume rather than mass.

We find that the median density of cold-mode gas is higher, by up to 1~dex, than that of hot-mode gas and that its current temperature is lower, by up to 2~dex, at least beyond $0.2R_\mathrm{vir}$. The hot-mode maximum past temperature is on average about an order of magnitude higher than the cold-mode maximum past temperature. The median pressure of the hot-mode gas only exceeds that of the cold mode by a factor of a few, but for the entropy the difference reaches 2.5~dex. For $R\ga R_{\rm vir}$ the gas metallicity in the cold mode is lower and has a much larger spread (four times larger at $R_\mathrm{vir}$) than in the hot mode. Cold-mode gas is flowing in at much higher velocities, by up to 150~$\rm km\, s^{-1}$, and dominates the accretion rate at all radii. Below, we will discuss these gas properties individually and in more detail.

\subsection{Density} \label{sec:dens}

The halo shown in Fig.~\ref{fig:halo} is being fed by dense, clumpy filaments as well as by cooling diffuse gas. The filaments are overdense for their radius, both inside and outside the halo. The top-left panel of Fig.~\ref{fig:haloradz2} shows that the overdensity of both hot- and cold-mode gas increases with decreasing radius, from $\sim 10$ at $10R_{\rm vir}$ to $\sim 10^2$ at $R_{\rm vir}$ and to $10^4$ at $0.1R_{\rm vir}$. The median density of cold-mode gas is higher by up to an order of magnitude than that of hot-mode gas for all radii $0.1R_\mathrm{vir}<R<4R_\mathrm{vir}$. The cold-mode gas densities exhibit a significant scatter of about 2~dex, as opposed to about 0.4~dex for hot-mode gas at $R_\mathrm{vir}$, which implies that the cold-mode gas is much clumpier. Beyond $4R_\mathrm{vir}$ the median hot-mode density becomes higher than for the cold mode, because there the hot-mode gas is associated with different haloes and/or large-scale filaments, which are also responsible for heating the gas.

\subsection{Temperature}

Hot gas, heated either by accretion shocks or by SN feedback, extends far beyond the virial radius (top-middle of Fig.~\ref{fig:halo}). Most of the volume is filled with hot gas. The location of cold gas overlaps with that of dense gas, so the temperature and density are anti-correlated. This anti-correlation is a result of the fact that the cooling time deceases with the gas density.

For $R\gtrsim0.2R_\mathrm{vir}$ the temperatures of the hot- and cold-mode gas do not vary strongly with radius (top middle panel of Fig.~\ref{fig:haloradz2}).  Note that this panel shows the \textit{current} temperature and not the maximum \textit{past} temperature. Gas accreted in the hot mode has a temperature $\sim 10^6$~K at $R>0.2R_\mathrm{vir}$, which is similar to the virial temperature. The median temperature of the hot-mode gas increases slightly from $\approx 2R_{\rm vir}$ to $\approx 0.2R_\mathrm{vir}$ because the hot gas is compressed as it falls in. Within $0.5R_{\rm vir}$ the scatter increases and around $0.2R_{\rm vir}$ the median temperature drops sharply to $\sim 10^4$~K. The dramatic decrease in the temperature of the hot-mode gas is a manifestation of the strong cooling flow that results when the gas has become sufficiently dense to radiate away its thermal energy within a dynamical time. The median temperature of cold-mode gas peaks at slightly below $10^5$~K around $2R_\mathrm{vir}$ and decreases to $\sim 10^4$~K at $0.1~R_\mathrm{vir}$. The peak in the temperature of the cold-mode gas is determined by the interplay between photo-heating by the UV background and radiative cooling. The temperature difference between the two accretion modes reaches a maximum of about 2~dex at $0.3R_\mathrm{vir}$ and vanishes  around $0.1R_\mathrm{vir}$.

\subsection{Maximum past temperature} 

The maximum past temperature (top-right panel of Fig.~\ref{fig:halo}) is by definition at least as high as the current temperature, but its spatial distribution correlates well with that of the current temperature. As shown by Fig.~\ref{fig:haloradz2}, the difference between maximum past temperature and current temperature is small at $R\ga R_\mathrm{vir}$, but increases towards smaller radii and becomes 1~dex for cold-mode gas and 2~dex for hot-mode gas at 0.1$R_\mathrm{vir}$.

While the temperature of the cold-mode gas decreases with decreasing radius, its maximum past temperature stays constant at $T_\mathrm{max}\approx10^5$~K. This value of $T_\mathrm{max}$ is reached around 2$R_\mathrm{vir}$ as a result of heating by the UV background. Both the current and the maximum past temperature of the hot-mode gas increase with decreasing radius for $R>0.3R_\mathrm{vir}$. The fact that $T_\mathrm{max}$ decreases below $0.3R_\mathrm{vir}$ shows that it is, on average, the colder part of the hot-mode gas that can reach these inner radii. If it were a random subset of all the hot-mode gas, then $T_\mathrm{max}$ would have stayed constant.

\subsection{Pressure}

As required by hydrostatic equilibrium, the gas pressure generally increases with decreasing radius (middle-left panels of Figs.~\ref{fig:halo} and \ref{fig:haloradz2}). However, the median pressure profile (Fig.~\ref{fig:haloradz2}) does show a dip around $0.2-0.3 R_{\rm vir}$ that reflects the sharp drop in the temperature profiles. Here catastrophic cooling leads to a strong cooling flow and thus a breakdown of hydrostatic equilibrium.

Comparing the pressure map with those of the density and temperature, the most striking difference is that the filaments become nearly invisible inside the virial radius, whereas they stood out in the density and temperature maps. However, beyond the virial radius the filaments do have a higher pressure than the diffuse gas. This suggests that pressure equilibrium is quickly established after the gas accretes onto the haloes. 

Fig.~\ref{fig:haloradz2} shows that the difference between the pressures of the hot- and cold-mode gas increases beyond $2R_\mathrm{vir}$. This is because at these large radii the hot-mode gas is associated with other haloes and/or large-scale filaments, while the cold-mode gas is intergalactic, so we do not expect them to be in pressure equilibrium.
Moving inwards from the virial radius, the median pressure difference increases until it reaches about an order of magnitude at $0.3R_{\rm vir}$. At smaller radii the pressures become nearly the same because the hot-mode gas cools down to the same temperature as the cold-mode gas. 

Although it takes some time to reach pressure equilibrium if the hot gas is suddenly heated, we expect the hot and cold gas to be approximately in equilibrium inside the halo, because a phase with a higher pressure will expand, lowering its pressure, and compressing the phase with the lower pressure, until equilibrium is reached. While the pressure distributions do overlap, there is still a significant difference between the two. This difference decreases somewhat with increasing resolution, because the cold gas reaches higher densities and thus higher pressures, as is shown in the appendix. From the example pressure map (Fig.~\ref{fig:halo}) we can see that the filaments inside the halo are in fact approximately in pressure equilibrium with the diffuse gas around them. At first sight this seems at odds with the fact that the median pressure profiles are different. However, the pressure map also reveals an asymmetry in the pressure inside the halo, with the gas to the left of the centre having a higher pressure than the gas to the right of the centre. Because there is also more hot-mode gas to the left, this leads to a pressure difference between the two modes when averaged over spherical shells, even though the two phases are locally in equilibrium. The asymmetry arises because the hot-mode gas is a space-filling gas and the flow has to converge towards the centre of the halo, which increases its pressure. The cold-mode gas is not space filling and therefore does not need to compress as much.

\subsection{Entropy}

We define the entropy as 
\begin{equation}
S\equiv \dfrac{P(\mu m_\mathrm{H})^{5/3}}{k_\mathrm{B}\rho^{5/3}},
\end{equation}
where $\mu$ is the mean molecular weight, $m_\mathrm{H}$ is the mass of a hydrogen atom, and $k_\mathrm{B}$ is Boltzmann's constant. Note that the entropy remains invariant for adiabatic processes. In the central panel of Fig.~\ref{fig:halo} we clearly see that the filaments have much lower entropies than the diffuse gas around them. This is expected for cold, dense gas.

For $R>R_\mathrm{vir}$ the median entropy of hot-mode gas is always higher than that of cold-mode gas (Fig.~\ref{fig:haloradz2}). While the entropy of the cold-mode gas decreases smoothly and strongly towards the centre of the halo, the entropy of the hot-mode gas decreases only slightly down to $0.2R_\mathrm{vir}$ after which it drops steeply. Cold-mode gas cools gradually, but hot-mode gas cannot cool until it reaches high enough densities, which results in a strong cooling flow.

\subsection{Metallicity} \label{sec:metal}

The middle-right panels of Figs.~\ref{fig:halo} and \ref{fig:haloradz2} show that the cold-mode streams have much lower metallicities than the diffuse, hot-mode gas, at least for $R\ga R_{\rm vir}$. The cold-mode gas also has a much larger
spread in metallicity, four orders of magnitude at $R_\mathrm{vir}$,
as opposed to only one order of magnitude for hot-mode gas. 

The gas in the filaments tends to have a lower metallicity, because most of it has never been close to a star-forming region, nor has it been affected by galactic winds, which tend to avoid the filaments \citep{Theuns2002}. The radial velocity image in the bottom-left panel of Fig.~\ref{fig:halo} confirms that the winds take the path of least resistance. The cold mode also includes dense clumps, which show a wide range of metallicities. If the density is high enough for embedded star formation to occur, then this can quickly enrich the entire clump. The enhanced metallicity will increase its cooling rate, making it even more likely to accrete in the cold mode (recall that our definition of the maximum past temperature ignores shocks in the ISM). On the other hand, clumps that have not formed stars remain metal-poor. The metallicity spread is thus caused by a combination of being shielded from winds driven by the central galaxy and exposure to internal star formation.
 
At $R_\mathrm{vir}$ the median metallicity of the gas is subsolar, $Z\sim 10^{-1}~Z_\odot$ for hot-mode gas and $Z\sim 10^{-2}~Z_\odot$ for cold-mode gas. However, we caution the reader that the median cold-mode metallicity is not converged with numerical resolution (see the Appendix) and could in fact be much lower. The metallicity increases towards the centre of the halo and this increase is steeper for cold-mode gas. The metallicity difference between the two modes disappears at $R\approx 0.5R_\mathrm{vir}$, but we find this radius to move inwards with increasing resolution (see the Appendix). 
Close to the centre the hot gas cools down and ongoing star formation in the disc enriches all the gas. The scatter in the metallicity decreases, especially for cold-mode gas, to $\sim0.7$~dex. 

As discussed in detail by \citet{Wiersma2009b}, there is no unique definition of metallicity in SPH. The metallicity that we assign to each particle is the ratio of the metal mass density and the total gas density at the position of the particle. These ``SPH-smoothed abundances'' were also used during the simulation for the calculation of the cooling rates. Instead of using SPH-smoothed metallicities, we could, however, also have chosen to compute the metallicity as the ratio of the metal mass and the total gas mass of each particle. Using these so-called particle metallicities would sharpen the metallicity gradients at the interfaces of different gas phases. Indeed, we find that using particle metallicities decreases the median metallicity of the metal-poor cold mode. For the hot mode the median particle metallicity is also lower than the median smoothed metallicity, but it increases with resolution, whereas the cold-mode particle metallicities decrease with resolution.

While high-metallicity gas may belong to either mode, gas with metallicity $\la 10^{-3}~Z_\odot$ is highly likely to be part of a cold flow. This conclusion is strengthened when we increase the resolution of the simulation or when we use particle rather than SPH-smoothed metallicities. Thus, a very low metallicity appears to be a robust way of identifying cold-mode gas.

\subsection{Radial velocity}

The radial peculiar velocity is calculated with respect to the halo centre after subtracting the peculiar velocity of the halo. The peculiar velocity of the halo is calculated by taking the mass-weighted average velocity of all the gas particles within 10 per cent of virial radius. Note that the Hubble flow is not included in the radial velocities shown. It is unimportant inside haloes, but is about a factor of two larger than the peculiar velocity at $10R_\mathrm{vir}$. 

The bottom-left panel of Fig.~\ref{fig:halo} shows that gas outside the haloes is, in general, moving towards the halo (i.e.\ it has a negative radial velocity). Within the virial radius, however, more than half of the projected area is covered by outflowing gas. These outflows are not only caused by SN feedback. In fact, simulations without feedback also show significant outflows (see Figs.~\ref{fig:halodiff} and \ref{fig:haloradz2diff}). A comparison with the pressure map shows that the outflows occur in the regions where the pressure is relatively high for its radius (by $0.1-0.4$~dex, see Figs.~\ref{fig:haloradflux} and \ref{fig:halomassflux}). The inflowing gas is associated with the dense, cold streams, but the regions of infall are broader than the cold filaments. Some of the hot-mode gas is also flowing in along with the cold-mode gas. These fast streams can penetrate the halo and feed the central disc. At the same time, some of the high temperature gas will expand, causing mild outflows in high pressure regions and these outflows are strengthened by SN-driven winds.

The hot-mode gas is falling in more slowly than the cold-mode gas or is even outflowing (bottom-left panel of Fig.~\ref{fig:haloradz2}). This is expected, because the gas converts its kinetic energy into thermal energy when it goes through an accretion shock and because a significant fraction of the hot-mode gas may have been affected by feedback. For hot-mode gas the median radial velocity is closest to zero between $0.3R_\mathrm{vir}<R<1R_\mathrm{vir}$. Most of it is inflowing at smaller radii, where the gas temperature drops dramatically, and also at larger radii. 

The cold-mode gas appears to accelerate to $-150$~km s$^{-1}$ towards $R_\mathrm{vir}$ (i.e.\ radial velocities becoming more negative) and to decelerate to $-30$ km s$^{-1}$ from $R_\mathrm{vir}$ towards the disc. We stress, however, that the behaviour of individual gas elements is likely to differ significantly from the median profiles. Individual, cold gas parcels will likely accelerate until they go through an accretion shock or are hit by an outflow, at which point the radial velocity may suddenly vanish or change sign. If this is more likely to happen at smaller radii, then the median profiles will show a smoothly decelerating inflow. Finally, observe that while there is almost no outflowing cold-mode gas around the virial radius, close to the central galaxy
($R\la 0.3R_\mathrm{vir}$) a significant fraction is outflowing.

\subsection{Accretion rate}

The appropriate definition of the accretion rate in an expanding Universe depends on the question of interest. Here we are interested in the mass growth of haloes in a comoving frame, where the haloes are defined using a criterion that would keep halo masses constant in time if there were no peculiar velocities. An example of such a halo definition is the spherical overdensity criterion, which we use here, because the virial radius is in that case defined as the radius within which the mean internal density is a fixed multiple of some fixed comoving density. 

The net amount of gas mass that is accreted per unit time through a spherical shell $S$ with comoving radius $x = R/a$, where $a$ is the expansion factor, is then given by the surface integral
\begin{eqnarray}
\dot{M}_{\rm gas}(x) &=& - \int_S a^3 \rho  \dot{x}\frac{dS}{a^2}\\
&=& - \int_S \rho v_{\rm rad} dS,
\end{eqnarray}
where $a^3\rho$ and $dS/a^2$ are a comoving density and a comoving area, respectively, and the radial peculiar velocity is $v_{\rm rad} \equiv a\dot{x}$. We evaluate this integral as follows,
\begin{equation}
\dot{M}_{\rm gas}(R) = - \sum_{R \le r_i < R+dR} \dfrac{m_{{\rm
gas},i}v_{{\rm rad},i}}{V_\mathrm{shell}}A_\mathrm{shell},
\label{eq:mdotgas}
\end{equation}
where
\begin{equation}
V_\mathrm{shell}=\dfrac{4\pi}{3}((R+dR)^3-R^3),
\end{equation}
\begin{equation}
A_\mathrm{shell}=4\pi(R+\dfrac{1}{2}dR)^2,
\end{equation}
$r_i$ is the radius of particle $i$, and $dR$ is the bin size. Note that a negative accretion rate corresponds to net outflow.

The mass flux map shown in the bottom-middle panel of Fig.~\ref{fig:halo} is computed per unit area for each pixel as $\Sigma_i m_{{\rm gas},i}v_{{\rm rad},i}/V_\mathrm{pix}$, where $V_\mathrm{pix}$ is the proper volume of the pixel. The absolute mass flux is highest in the dense filaments and in the other galaxies outside $R_\mathrm{vir}$, because they contain a lot of mass and have high inflow velocities.

The gas accretion rate $\dot{M}_{\rm gas}(R/R_{\rm vir})$ computed using equation~(\ref{eq:mdotgas}) is shown as the black curve in the bottom-middle panel of Fig.~\ref{fig:haloradz2}. The accretion rate is averaged over all haloes in the mass bin we are considering here ($10^{11.5}$~M$_\odot<M_\mathrm{halo}<10^{12.5}$~M$_\odot$). Similarly, the red and blue curves are computed by including only hot- and cold-mode particles, respectively. The accretion rate is positive at all radii, indicating net accretion for both modes. The inflow rate is higher for the cold mode even though the hot-mode gas dominates the mass budget around the virial radius (see Section~\ref{sec:hot}). The hot-mode gas accretion rate is a combination of the density and the radial velocity of the hot-mode gas, as well as the amount of mass in the hot mode. Gas belonging to the cold mode at $R>R_\mathrm{vir}$ may later become part of the hot mode after it has reached $R<R_\mathrm{vir}$.

The extended halo is not in a steady state, because the accretion rate varies with radius. Moving inwards from $10 R_{\rm vir}$ to $R_{\rm vir}$, the net rate of infall drops by about an order of magnitude. This implies that the (extended) halo is growing: the flux of mass that enters a shell from larger radii exceeds the flux of mass that
leaves the same shell in the direction of the halo centre. This sharp drop in the accretion rate with decreasing radius is in part due to the fact that some of the gas at $R > R_{\rm vir}$ is falling towards other haloes that trace the same large-scale structure.

Within the virial radius the rate of infall of all gas and of cold-mode gas continues to drop with decreasing radius, but the gradient becomes much less steep ($d\ln \dot{M}/d\ln R \approx 0.4$), indicating that the cold streams are efficient in transporting mass to the central galaxy. 
For the hot mode the accretion rate only flattens at $R\lesssim 0.4R_\mathrm{vir}$ around the onset of catastrophic cooling. Hence, once the hot-mode gas reaches small enough radii, its density becomes sufficiently high for cooling to become efficient, and the hot-mode accretion becomes efficient too. However, even at 0.1$R_\mathrm{vir}$ its accretion rate is still much lower than that of cold-mode gas.

\subsection{Hot fraction} \label{sec:hot}

Even though the average hot fraction, i.e.\ the mean fraction of the gas mass that has a maximum past temperature greater than $10^{5.5}$~K, of the halo in the image is close to 0.5, few of the pixels actually have this value. For most pixels $f_\mathrm{hot}$ is either close to one or zero (bottom-right panel of Fig.~\ref{fig:halo}), confirming the bimodal nature of the accretion.

The bottom-right panel of Fig.~\ref{fig:haloradz2} shows that the hot fraction peaks around the virial radius, where it is about 70 per cent. Although the hot fraction decreases beyond the virial radius, it is still 30 per cent around $10 R_{\rm vir}$. The hot-mode gas at very large radii is associated with other haloes and/or large-scale filaments. Within the halo the hot fraction decreases from 70 per cent at $R_\mathrm{vir}$ to 35 per cent at $0.1R_\mathrm{vir}$. While hot-mode accretion dominates the growth of haloes, most of the hot-mode gas does not reach the centre. Cold-mode accretion thus dominates the growth of galaxies.

\section{Dependence on halo mass}
\label{sec:mass}

\begin{figure*}
\center
\includegraphics[scale=0.65]{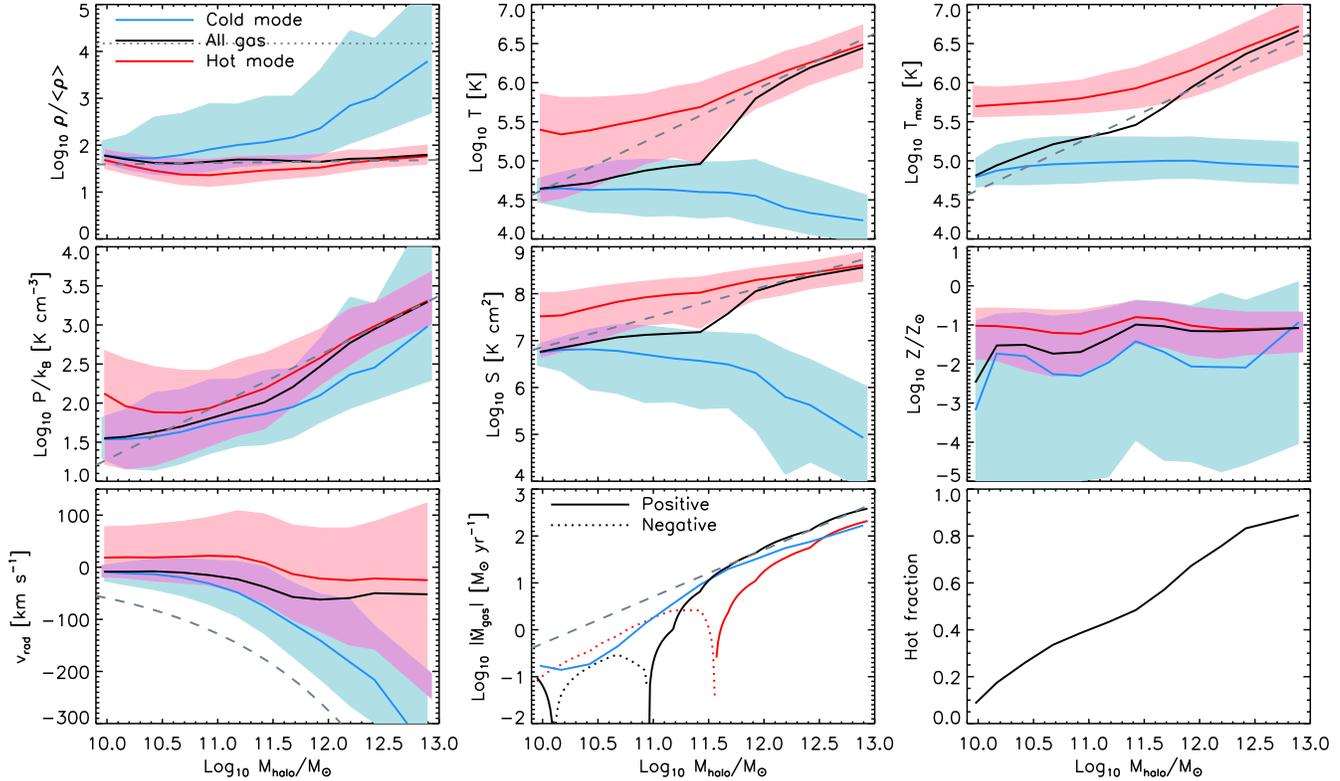}
\caption {\label{fig:halomassz2} Properties of gas at $0.8R_\mathrm{vir}<R<R_\mathrm{vir}$ at $z=2$ as a function of halo mass for all (black curves), hot-mode (red curves), and cold-mode (blue curves) gas. Results are shown for the simulation \emph{REF\_L050N512}. Shaded regions show values within the 16th and 84th percentiles, i.e., the $\pm 1\sigma$ scatter around the median. From the top-left to the bottom-right, the different panels show the mass-weighted median gas overdensity, temperature, maximum past temperature, pressure, entropy, metallicity, radial peculiar velocity, the mean accretion rate, and the mean mass fraction of hot-mode gas, respectively. The horizontal, dotted line in the first panel indicates the threshold for star formation ($n_{\rm H} = 0.1~{\rm cm}^{-3}$). The dashed, grey curves show analytic estimates from virial arguments.}
\end{figure*}

In Fig.~\ref{fig:halomassz2} we plot the same properties as in Fig.~\ref{fig:haloradz2} as a function of halo mass for gas at radii $0.8R_\mathrm{vir}<R<R_\mathrm{vir}$, where differences between hot- and cold-mode gas are large. Grey, dashed lines show analytic estimates and are discussed below. The dotted, grey line in the top-left panel indicates the star formation threshold, i.e.\ $n_{\rm H}=0.1~{\rm cm}^{-3}$. The differences between the density and temperature of the hot- and cold-mode gas increase with the mass of the halo. The average temperature, maximum past temperature, pressure, entropy, metallicity, absolute radial peculiar velocity, absolute accretion rate, and the hot fraction all increase with halo mass.

We can compare the gas overdensity at the virial radius to the density that we would expect if baryons were to trace the dark matter, $\rho_\mathrm{vir}$. We assume an NFW profile \citep{Navarro1996}, take the mean internal density relative to the critical density at redshift $z$, $\Delta_c\langle\rho\rangle$, from spherical collapse calculations \citep{Bryan1998} and the halo mass-concentration relation from \citet{Duffy2008} and calculate the mean overdensity at $R_\mathrm{vir}$. This is plotted as the dashed, grey line in the top-left panel of Fig.~\ref{fig:halomassz2}. It varies very weakly with halo mass, because the concentration depends on halo mass, but this is invisible on the scale of the plot. For all halo masses the median density is indeed close to this analytic estimate. While the same is true for the hot-mode gas, for high-mass haloes
($M_\mathrm{halo}\gtrsim10^{12}$~M$_\odot$) the median density of cold-mode gas is significantly higher than the estimated density and the difference reaches two orders of magnitude for $M_\mathrm{halo}\sim10^{13}$~M$_\odot$. A significant fraction of the cold-mode gas in these most massive haloes is star forming and hence part of the ISM of satellite galaxies. The fact that cold-mode gas becomes denser and thus clumpier with halo mass could have important consequences for the formation of clumpy galaxies at high redshift \citep{Dekel2009b, Agertz2009, Ceverino2010}.

The blue curve in the top-middle panel shows that the median temperature of the cold-mode gas at $R_\mathrm{vir}$ decreases slightly with halo mass, from 40,000~K to 15,000~K. This reflects the increase in the median density of cold-mode gas with halo mass, which results in shorter cooling times. The median temperature of the hot-mode gas
increases with halo mass and is approximately equal
to the virial temperature for 
$M_\mathrm{halo}\gtrsim10^{11.5}$~M$_\odot$. The virial temperature is plotted as the dashed, grey line and is given by 
\begin{eqnarray} \label{eqn:virialtemperature}
T_\mathrm{vir}&=&\left(\dfrac{G^2H_0^2\Omega_\mathrm{m}\Delta_c}{54}\right)^{1/3}\dfrac{\mu m_\mathrm{H}}{k_B}M_{\rm halo}^{2/3}(1+z),\\
&\approx& 9.1\times10^5~{\rm K} \left (\dfrac{M_{\rm halo}}{10^{12}~{\rm M}_\odot}\right )^{2/3} \left (\dfrac{1+z}{3}\right ),
\end{eqnarray}
where $G$ is the gravitational constant, $H_0$ the Hubble constant and $\mu$ is assumed to be equal to 0.59.

While much of the gas accreted onto low-mass haloes in the hot mode has a temperature smaller than $10^{5.5}$~K and has therefore already cooled down substantially\footnote{Note that for haloes with $T_{\rm vir} \la 10^{5.5}$~K the median hot-mode temperature is affected by the requirement $T_{\rm max} > 10^{5.5}$~K (our definition of hot-mode gas).}, there is very little overlap in
the current temperatures of gas accreted in the two modes for haloes with $T_\mathrm{vir} \ga 10^6~$K. Because the cooling rates decrease with temperature for $T > 10^{5.5}~$K \citep[e.g.][]{Wiersma2009a}, most of the hot-mode gas in haloes with higher temperatures stays hot. For the same reason, the median temperature of all gas rises sharply at $M\approx10^{11.5}$~M$_\odot$ ($T_{\rm vir} \approx 10^{5.5}$~K) and is roughly equal to $T_\mathrm{vir}$ for $M_\mathrm{halo}>10^{12}$~M$_\odot$.

The top-right panel shows that the median maximum past temperature of gas at the virial radius is close to the virial temperature, which is again shown as the grey dashed curve, for the full range of halo masses shown. Some of the gas does, however, have a maximum past temperatures that differs strongly from the virial temperature. The largest difference is found for cold-mode gas in high-mass haloes. Because of its high density, its cooling time is short and the gas does not shock to the virial temperature. The maximum past temperature of gas accreted in the cold mode is close to $10^5$~K for all halo masses. For $M_\mathrm{halo} < 10^{10.5}$~M$_\odot$ this temperature is higher than the virial temperature. The gas in low-mass haloes has not been heated to its maximum temperature by a virial shock, but by the UV background radiation or by shocks from galactic winds. Heating by the UV background is the dominant process, because simulations without supernova feedback show the same result (see Fig.~\ref{fig:haloradz2diff}).
The maximum past temperature of hot-mode gas follows the virial temperature closely for high-mass haloes. For $T_\mathrm{vir}<10^{5.5}$~K ($M_\mathrm{halo} \la 10^{11.5}$~M$_\odot$) the maximum past temperature of the hot-mode gas remains approximately constant, at around $10^{5.7}$~K, because of our definition of hot-mode gas ($T_\mathrm{max}\ge10^{5.5}$~K).

The pressure of the gas increases roughly as $M_{\rm halo}^{2/3}$ (middle-left panel). We can estimate the pressure at the virial radius from the virial temperature and the density at the virial radius.
\begin{equation}
\frac{P_\mathrm{vir}}{k_{\rm B}} = \dfrac{T_\mathrm{vir}\rho_\mathrm{vir}}{\mu m_\mathrm{H}},
\end{equation}
where $\mu$ is assumed to be equal to 0.59. This pressure is shown by the dashed, grey line. The actual pressure is very close to this simple estimate. It scales with mass as the virial temperature because the density at the virial radius is nearly independent of the mass. For all halo masses the median pressure of the gas accreted in the hot-mode is about a factor of two higher than the median pressure of the cold-mode gas. 

The central panel shows that the entropy difference between hot- and cold-mode gas increases with halo mass, because the entropy of hot-mode gas increases with halo mass, whereas the cold-mode entropy decreases. The hot-mode gas follows the slope of the relation expected from virial arguments, 
\begin{equation}
S_\mathrm{vir}=\dfrac{P_\mathrm{vir}(\mu m_\mathrm{H})^{5/3}}{k_\mathrm{B}\rho_\mathrm{vir}^{5/3}},
\end{equation}
where $\mu$ is assumed to be equal to 0.59. $S_\mathrm{vir}$ is shown as the dashed, grey line.

The middle-right panel shows that the median gas metallicity at the virial radius increases from $\sim 10^{-2}~Z_\odot$ for $M_{\rm halo} \sim 10^{10}~{\rm M}_\odot$ to $\sim
10^{-1}~Z_\odot$ for $10^{13}~{\rm M}_\odot$. This increase reflects the increased fraction of hot-mode gas (see the bottom-right panel) and an increase in the median metallicity of the cold-mode gas, which is probably due to the fact that a greater fraction of the gas resides in the ISM of satellite galaxies for more massive haloes (see the top-left panel). The scatter in the metallicity of the cold-mode gas is always very large. The hot-mode gas has a median metallicity $\sim 10^{-1}~Z_\odot$ for all halo masses, which is similar to the predicted metallicity of the warm-hot intergalactic medium \citep{Wiersma2011}.  

The black curve in the bottom-left panel shows that for all halo masses more mass is falling into the halo than is flowing out. The radial velocity distributions are, however, very broad. A substantial fraction of the hot-mode gas, more than half for $M_{\rm halo}< 10^{11.5}~{\rm M}_\odot$, is outflowing at $R_\mathrm{vir}$. Cold-mode gas is predominantly inflowing for all masses, but the fraction of outflowing gas becomes significant for $M_{\rm halo}< 10^{11.5}~{\rm M}_\odot$. 

As expected, the gas at the virial radius falls in faster for higher-mass haloes and the absolute velocities are larger for cold-mode gas. 
We can compare the radial peculiar velocity to the escape velocity,
\begin{eqnarray}\label{eqn:esc}
v_\mathrm{esc} &=& \sqrt{\dfrac{2GM_{\rm halo}}{R_\mathrm{vir}}},\\
&\approx & 275~{\rm km}\,{\rm s}^{-1} \left (\frac{M_{\rm halo}}{10^{12}~{\rm M}_\odot}\right )^{1/3} \left (\frac{1+z}{3}\right )^{1/2}, 
\end{eqnarray}
where we assumed a matter-dominated Universe and used
\begin{eqnarray}
R_\mathrm{vir} &=& \left(\dfrac{2GM}{H_0^2\Omega_\mathrm{m}\Delta_c}\right)^{1/3}\dfrac{1}{1+z},\\
&\approx & 114~{\rm kpc} \left( \frac{M_{\rm halo}}{10^{12}~{\rm M}_\odot}\right )^{1/3} \left (\frac{1+z}{3}\right )^{-1}.
\end{eqnarray}
We show $-v_\mathrm{esc}$ by the dashed, grey curve. We only expect the gas to have a velocity close to this estimate if it fell in freely from very large distances and if the Hubble expansion, which damps peculiar velocities, were unimportant. However, we do expect the scaling with mass to be more generally applicable. 
For the cold mode the trend with halo mass is indeed well reproduced by the escape velocity.

\begin{figure*}
\center
\includegraphics[scale=0.65]{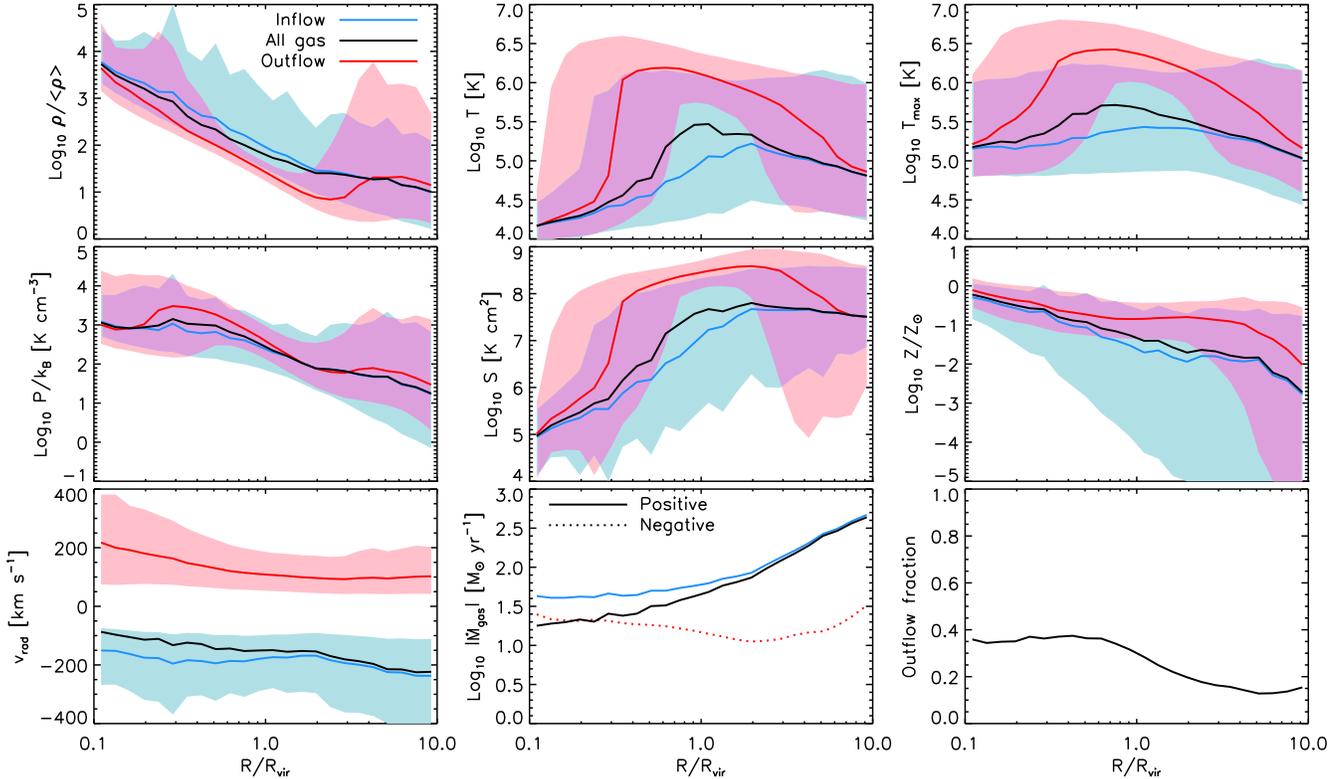}
\caption {\label{fig:haloradflux}  Properties of gas in haloes with $10^{11.5}$~M$_\odot<M_\mathrm{halo}<10^{12.5}$~M$_\odot$ at $z=2$ as a function of radius for all (black curves), outflowing (red curves), and inflowing (blue curves) gas. Results are shown for the simulation \emph{REF\_L050N512}. Shaded regions show values within the 16th and 84th percentiles, i.e., the $\pm 1\sigma$ scatter around the median. From the top-left to the bottom-right, the different panels show the mass flux-weighted median gas overdensity, temperature, maximum past temperature, pressure, entropy, metallicity, radial peculiar velocity, the mean accretion rate, and the mean mass fraction of outflowing gas, respectively. Most of the trends with radius for inflowing and outflowing gas are similar to those for cold-mode and hot-mode gas, respectively, as shown in Fig.~\ref{fig:haloradz2}.}
\end{figure*}
\begin{figure*}
\center
\includegraphics[scale=0.65]{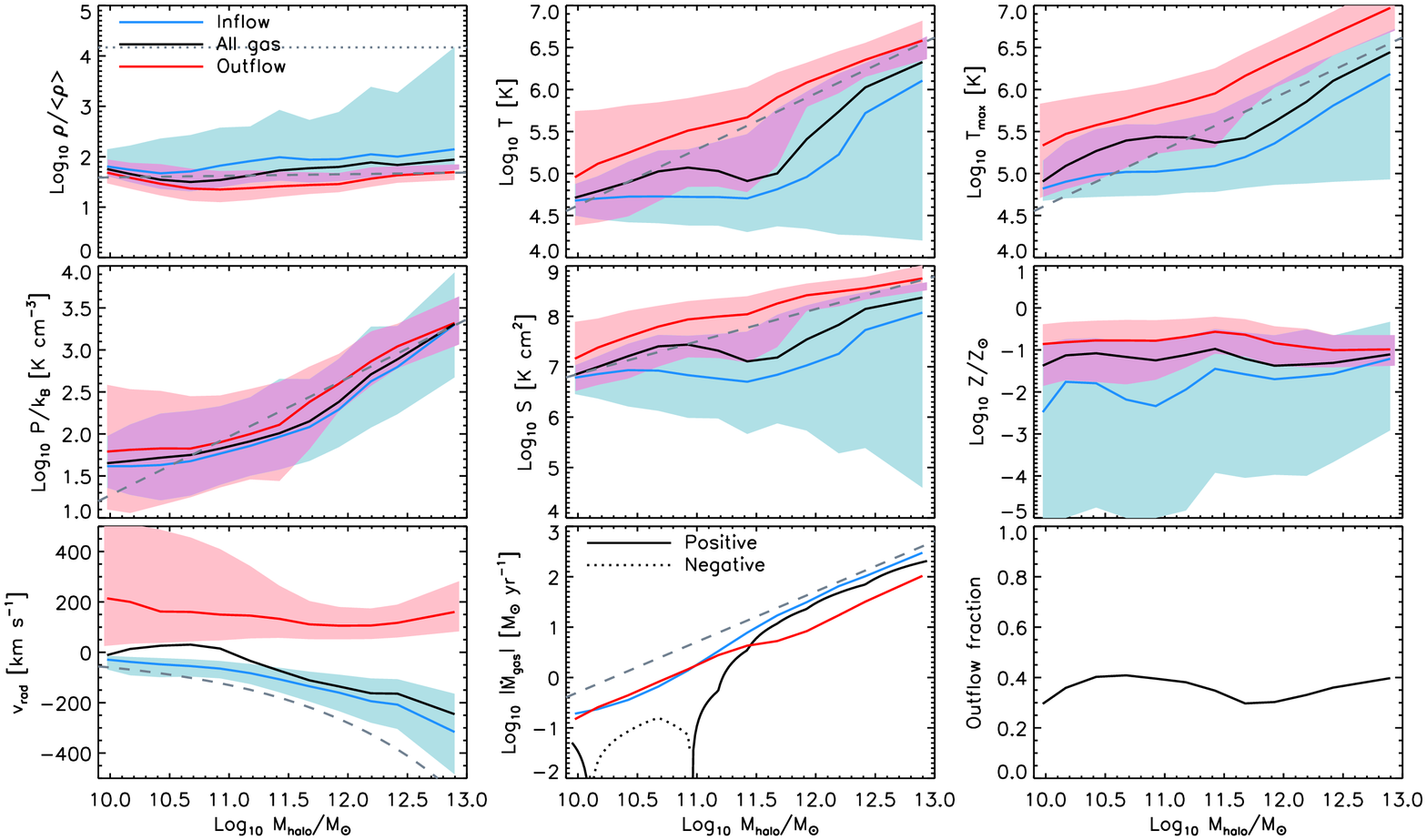}
\caption {\label{fig:halomassflux} Properties of gas at $0.8R_\mathrm{vir}<R<R_\mathrm{vir}$ at $z=2$ as a function of halo mass for all (black curves), outflowing (red curves), and inflowing (blue curves) gas. Results are shown for the simulation \emph{REF\_L050N512}. Shaded regions show values within the 16th and 84th percentile, i.e., the $\pm 1\sigma$ scatter around the median. From the top-left to the bottom-right, the different panels show the mass flux-weighted median gas overdensity, temperature, maximum past temperature, pressure, entropy, metallicity, radial peculiar velocity, the mean accretion rate, and the mean mass fraction of outflowing gas, respectively. The horizontal, dotted line in the first panel indicates the threshold for star formation ($n_{\rm H}=0.1~{\rm cm}^{-3}$). The dashed, grey curves show analytic estimates from virial arguments. Most of the trends with mass for inflowing and outflowing gas are similar to those for cold-mode and hot-mode gas, respectively, as shown in Fig.~\ref{fig:halomassz2}.}
\end{figure*}

On average, for halo masses above $10^{11}$~M$_\odot$, the net accretion rate is positive, which means that more mass is flowing in than is flowing out (bottom-middle panel). Therefore, unsurprisingly, the haloes are growing. For haloes with $10^{10}$~M$_\odot<M_\mathrm{halo}<10^{11}$~M$_\odot$ the mean accretion rate is negative (indicating net outflow), but small ($\sim 0.1$~M$_\odot \,$yr$^{-1}$). The mean accretion rate of cold-mode gas is positive for all halo masses, but for hot-mode gas there is net outflow for $M_\mathrm{halo}<10^{11.5}$~M$_\odot$. Although these haloes are losing gas that is currently hot-mode, their hot-mode gas reservoir may still be increasing if cold-mode gas is converted into hot-mode gas. For higher-mass haloes, the hot-mode accretion rate is also positive and it increases approximately linearly with halo mass. This is the regime where the implemented supernova feedback is not strong enough to blow gas out of the halo. This transition mass is increased by more than an order of magnitude when more effective supernova feedback or AGN feedback is included (not shown). For $M_\mathrm{halo}>10^{12.5}$~M$_\odot$ the hot-mode inflow rate is slightly stronger than the cold-mode inflow rate. 

The grey, dashed curve indicates the accretion rate a halo with a baryon fraction $\Omega_{\rm b}/\Omega_{\rm m}$ would need to have to grow to its current baryonic mass in a time equal to the age of the Universe at $z=2$,
\begin{equation}
\dot{M}=\dfrac{\Omega_\mathrm{b}M_\mathrm{halo}}{\Omega_\mathrm{m}t_\mathrm{Universe}}.
\end{equation}
Comparing this analytic estimate with the actual mean accretion rate, we see that they are equal for $M_\mathrm{halo}>10^{11.5}$~M$_\odot$, indicating that these haloes are in a regime of efficient growth. For lower-mass haloes, the infall rates are much lower, indicating that the growth of these haloes has halted, or that their baryon fractions are much smaller than $\Omega_{\rm b}/\Omega_{\rm m}$.

The bottom-right panel of Fig.~\ref{fig:halomassz2} shows that, at the virial radius, hot-mode gas dominates the gas mass for high-mass haloes. The hot fraction at $R_\mathrm{vir}$ increases from 10 per cent in haloes of $\sim 10^{10}$~M$_\odot$ to 90 per cent for $M_\mathrm{halo}\sim 10^{13}$~M$_\odot$. Note that for haloes with $M_\mathrm{halo}<10^{11.3}$~M$_\odot$ the virial temperatures are lower than our adopted threshold for hot-mode gas. The hot fraction would have been much lower without supernova feedback ($f_\mathrm{hot}<5$~per cent for $M_\mathrm{halo}<10^{10.5}$~M$_\odot$).

\section{Inflow and outflow} \label{sec:inout}

Figs.~\ref{fig:haloradflux} and~\ref{fig:halomassflux} show the physical properties of the gas, weighted by the radial mass flux, for all, inflowing, and outflowing gas (black, blue, and red curves, respectively). Except for the last two panels, the curves indicate the medians, i.e.\ half the mass flux is due to gas above the curves. Similarly, the shaded regions indicate the 16th and 84th percentiles. Note that we do not plot this separately for hot- and cold-mode gas. The differences between the blue and red curves arise purely from the different radial peculiar velocity directions.

We can immediately see that separating gas according to its radial velocity direction yields similar results as when the gas is separated according to its maximum past temperature (compare to Figs.~\ref{fig:haloradz2} and~\ref{fig:halomassz2}). Like cold-mode gas, inflowing gas has, on average, a higher density, a lower temperature, a lower entropy, and a lower metallicity than outflowing gas. 

Comparing Figs.~\ref{fig:haloradz2} and~\ref{fig:haloradflux}, we notice that the differences in density, temperature, pressure, entropy, and metallicity between in- and outflowing gas tend to be slightly smaller than between cold- and hot-mode gas. This is particularly true outside the haloes (at $R>3R_\mathrm{vir}$), where there is a clear upturn in the density and pressure of outflowing gas that is accompanied by a marked decrease in the temperature. These features are due to gas that is flowing towards other haloes and/or large-scale filaments. Although such gas is outflowing from the perspective of the selected halo, it is actually infalling gas and hence more likely to be cold-mode.

Unsurprisingly, the radial peculiar velocities (bottom-left panels) are clearly very different when we separate the gas into in- and outflowing components than when we divide it into cold and hot modes. Low values of the radial velocity are avoided because the plot is weighted by the mass flux. While the radial velocity of cold-mode gas decreases from the virial radius towards the centre, the mass flux-weighted median radial velocity of inflowing gas is roughly constant. Within the haloes, the mass flux-weighted median radial velocity of outflowing gas decreases with radius. 
 
Both the inflow and the outflow mass flux (bottom-middle panel of Fig.~\ref{fig:haloradflux}) are approximately constant inside $0.7 R_{\rm vir}$, which implies that the fraction of the
gas that is outflowing is also constant (last panel of Fig.~\ref{fig:haloradflux}). A mass flux that is independent of radius implies efficient mass transport, as the same amount of mass passes through each shell per unit of time. The inflowing mass flux decreases from 10$R_\mathrm{vir}$ to $\sim R_\mathrm{vir}$ because the overdensity of the region is increasing and because some of the gas that is infalling at distances $\gg R_{\mathrm{vir}}$ is falling towards neighbouring haloes. At $R\gtrsim R_\mathrm{vir}$ the outflowing mass flux decreases somewhat, indicating that the transportation of outflowing material slows down and that the galactic winds are becoming less efficient. This can also be seen by the drop in outflow fraction around the virial radius in the last panel of Fig.~\ref{fig:haloradflux}. (The small decrease in outflow fraction below $10^{10.5}$~M$_\odot$ is a resolution effect.) The outflowing mass flux increases again at large radii, because the hot-mode gas is falling towards unrelated haloes.

Comparing Figs.~\ref{fig:halomassz2} and~\ref{fig:halomassflux}, we see again that, to first order, inflowing and outflowing gas behave similarly as cold-mode and hot-mode gas, respectively. There are, however, some clear differences, although we do need to keep in mind that some are due to the fact that Fig.~\ref{fig:halomassz2} is mass-weighted while Fig.~\ref{fig:halomassflux} is mass flux-weighted. The mass flux-weighted median temperature of outflowing gas is always close to the virial temperature. For $M_\mathrm{halo}\la 10^{11}$~M$_\odot$ this is much lower than the median temperature of hot-mode gas, but that merely reflects the fact that for these haloes the virial temperature is lower than the value of $10^{5.5}~$K that we use to separate the cold and hot modes. The mass flux-weighted maximum past temperature is about 0.5~dex higher than $T_\mathrm{vir}$.

Another clear difference is visible at the high-mass end ($M_\mathrm{halo}\ga 10^{12.5}$~M$_\odot$). While the cold-mode density increases rapidly with mass, the overdensity of infalling gas remains $\sim10^2$, and while the cold-mode temperature remains $\sim 10^4~$K, the temperature of infalling gas increases with halo mass. Both of these differences can be explained by noting that, around the virial radius, hot-mode gas accounts for a greater fraction of the infall in higher mass haloes (see the bottom-middle panel of Fig.~\ref{fig:halomassz2}). 

As was the case for the cold-mode gas, the radial peculiar velocity of infalling gas scales like the escape velocity (bottom-middle panel). Interestingly, although the mass flux-weighted median outflowing velocity is almost independent of halo mass, the high-velocity tail is much more prominent for low-mass haloes. Because the potential wells in these haloes are shallow and because the gas pressure is lower, the outflows are not slowed down as much before they reach the virial radius. The flux-weighted outflow velocities are larger than the inflow velocities for $M_\mathrm{halo}<10^{11.5}$~M$_\odot$, whereas the opposite is the case for higher-mass haloes.

Finally, the last panel of Fig.~\ref{fig:halomassflux} shows that the fraction of the gas that is outflowing around $R_{\rm vir}$ is relatively stable at about 30--40 per cent. Although the accretion rate is negative for $10^{10}$~M$_\odot<M_\mathrm{halo}<10^{11}$~M$_\odot$, which indicates net outflow, less than half of the gas is outflowing.

\section{Effect of metal-line cooling and outflows driven by supernovae and AGN} \label{sec:SNAGN}

\begin{figure*}
\center
\includegraphics[scale=0.46]{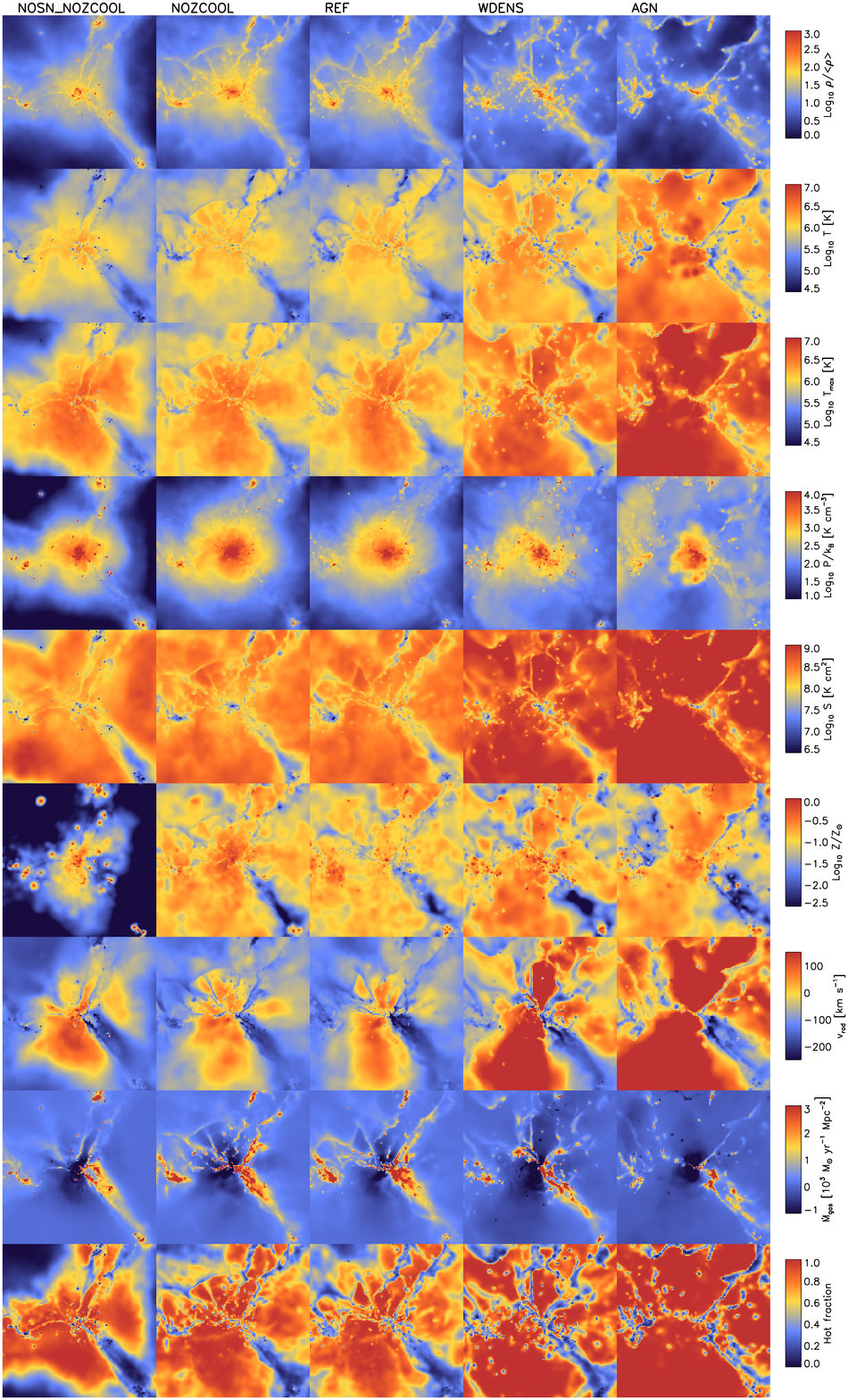}
\caption {\label{fig:halodiff} As Fig.~\ref{fig:halo}, but comparing the different simulations listed in Table~\ref{tab:owls}. The images show a cubic 1~$h^{-1}$~comoving Mpc region centred on a halo of $M_\mathrm{halo}\approx10^{12}$~M$_\odot$ at $z=2$ taken from the \emph{L025N512} simulations. From left to right, the different columns show the simulation with neither SN feedback nor metal-line cooling, the simulation without metal-line cooling, the reference simulation (which includes SN feedback and metal-line cooling), the simulation for which the wind speed scales with the effective sound speed of the ISM, and the simulation with AGN feedback. Note that the images shown in the middle column are identical to those shown in Fig.~\ref{fig:halo}.} 
\end{figure*}
\begin{figure*}
\center
\includegraphics[scale=0.65]{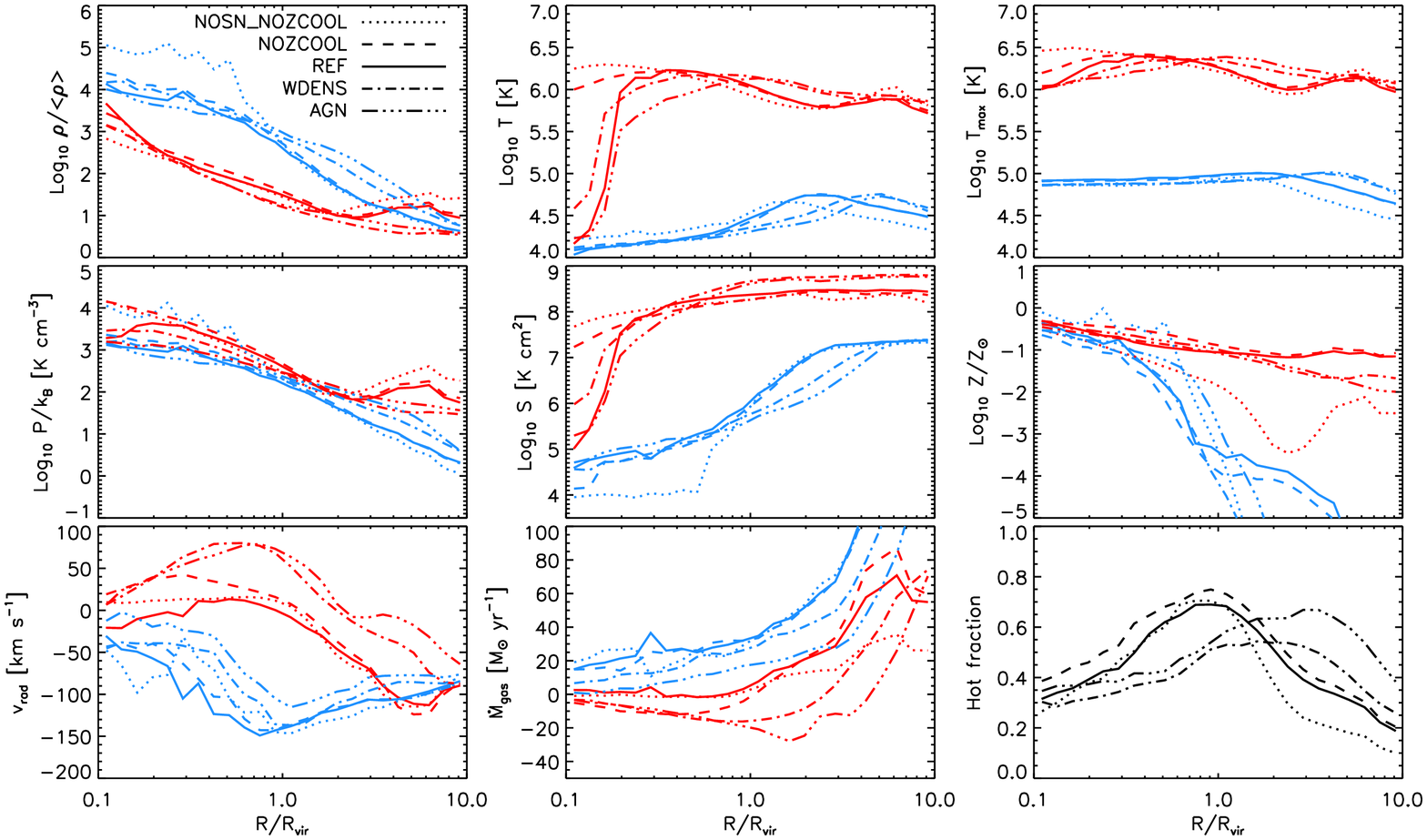}
\caption {\label{fig:haloradz2diff} Properties of gas in haloes with $10^{11.5}$~M$_\odot<M_\mathrm{halo}<10^{12.5}$~M$_\odot$ at $z=2$ as a function of radius for hot-mode (red curves) and cold-mode (blue curves) gas for the five different \emph{L025N112} simulations listed in Table~\ref{tab:owls}, as indicated in the legend.
From the top-left to the bottom-right, the different panels show the mass-weighted median gas overdensity, temperature, maximum past temperature, pressure, entropy, metallicity, radial peculiar velocity, the mean accretion rate, and the mean mass fraction of hot-mode gas, respectively.}
\end{figure*}

Fig.~\ref{fig:halodiff} shows images of the same $10^{12}$~M$_\odot$ halo as Fig.~\ref{fig:halo} for five different high-resolution (\emph{L025N512}) simulations at $z=2$. Each row shows a different property, in the same order as the panels in the previous figures. Different columns show different simulations, with the strength of galactic winds increasing from left to right (although the winds are somewhat stronger in \emph{NOZCOOL} than in \emph{REF}). The first column shows the simulation without SN feedback and without metal-line cooling. The second column shows the simulation with SN feedback, but without metal-line cooling. The third column shows our reference simulation, which includes both SN feedback and metal-line cooling. The fourth column shows the simulation with density-dependent SN feedback, which is more effective at creating galactic winds for this halo mass. The last column shows the simulation that includes both SN and AGN feedback.

The images show substantial and systematic differences. More efficient feedback results in lower densities and higher temperatures of diffuse gas and in the case of AGN feedback, even some of the cold filaments are partially destroyed. Although the images reveal some striking differences, Fig.~\ref{fig:haloradz2diff} shows that the trends in the profiles of the gas properties, including the differences between hot and cold modes, are very similar in the different simulations. This is partly because the profiles are mass-weighted, whereas the images are only mass-weighted along the projected dimension. We would have seen larger differences if we had shown volume-weighted profiles, because the low-density, high-temperature regions, which are most affected by the outflows, carry very little mass, but dominate the volume. Although the conclusions of the previous sections are to first order independent of the particular simulation that we use, there are some interesting and clear differences, which we shall discuss below.

We can isolate the effect of turning off SN-driven outflows by comparing models \emph{NOSN\_NOZCOOL} and \emph{NOZCOOL}. Without galactic winds, the cold-mode densities and pressures are about an order of magnitude higher within $0.5 R_{\rm vir}$. On the other hand, turning off winds decreases the density of hot-mode gas by nearly the same factor around $0.1 R_{\rm vir}$. Galactic winds thus limit the build up of cold-mode gas in the halo center, which they accomplish in part by converting cold-mode gas into hot-mode gas (i.e.\ by shock-heating cold-mode gas to temperatures above $10^{5.5}~$K). We can see that this must be the case by noting that the hot-mode accretion rate is negative inside the haloes for model \emph{NOZCOOL}. The absence of SN-driven outflows also has a large effect on the distribution of metals. Without winds, the metallicities of both hot- and cold-mode gas outside the halo are much lower, because there is no mechanism to transport metals to large distances. On the other hand, the metallicity of the cold-mode gas is higher within the halo, because the star formation rates, and thus the rates of metal production, are higher. This suggests that the metals in cold-mode gas are associated with locally formed stars, e.g.\ infalling companion galaxies.

Increasing the efficiency of the feedback, as in models \emph{WDENS} and particularly \emph{AGN}, the cold-mode median radial velocity becomes less negative and the accretion rate of cold-mode gas inside haloes decreases. At the same time, the radial velocity and the outflow rate of the hot-mode increase. In other words, more efficient feedback hinders the inflow rate of cold-mode gas and boosts the outflows of hot-mode gas. The differences are particularly large outside the haloes. Whereas the moderate feedback implemented in model \emph{REF} predicts net infall of hot-mode gas, the net accretion rate of hot-mode gas is negative out to about $4R_{\rm vir}$ when AGN feedback is included. Beyond the virial radius, stronger winds substantially increase the mass fraction of hot-mode gas. This comes at the expense of the hot-mode gas inside the haloes, which decreases if the feedback is more efficient, at least for $0.2 R_{\rm vir} < R < R_{\rm vir}$. 

The effect of metal-line cooling can be isolated by comparing models \emph{NOZCOOL} and \emph{REF}. Without metal-line cooling, the cooling times are much longer. Consequently, the median temperature of the hot-mode gas remains above $10^6$~K (at least for $R> 0.1R_\mathrm{vir}$), whereas it suddenly drops to below $10^5$~K around $0.2 R_{\rm vir}$ when metal-line cooling is included. Thus, the catastrophic cooling flow of the diffuse, hot component in the inner haloes is due to metals. Indeed, while the median hot-mode radial peculiar velocity within $0.2 R_{\rm vir}$ is positive without metal-line cooling, it becomes negative (i.e.\ infalling) when metal-line cooling is included.

\section{Evolution: Milky Way-sized haloes at $z=0$} \label{sec:z0}

\begin{figure*}
\center
\includegraphics[scale=0.65]{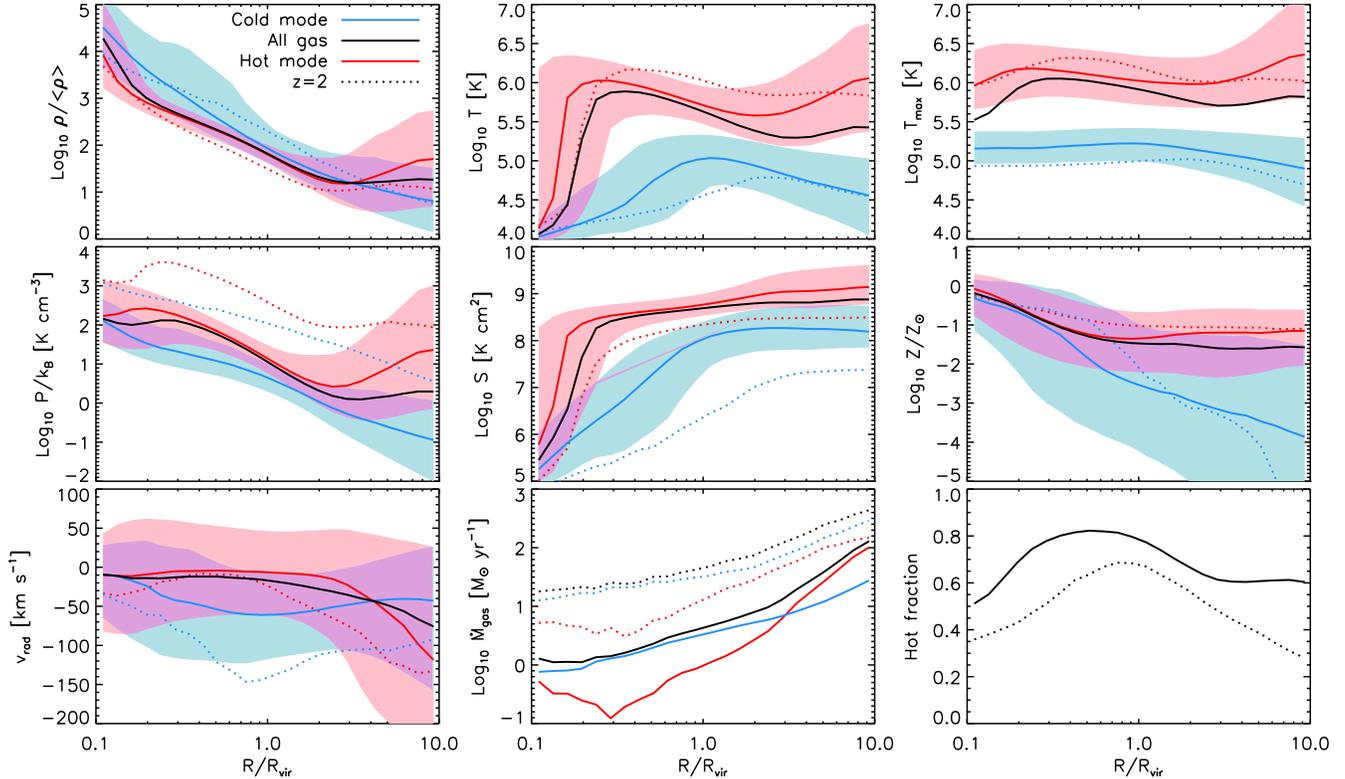}
\caption {\label{fig:haloradz0} Properties of gas in haloes with $10^{11.5}$~M$_\odot<M_\mathrm{halo}<10^{12.5}$~M$_\odot$ at $z=0$ as a function of radius for all (black curves), hot-mode (red curves), and cold-mode (blue curves) gas. Results are shown for the simulation \emph{REF\_L050N512}. Most $z=2$ curves from Fig.~\ref{fig:haloradz2} have also been included, for comparison, as dotted curves. Shaded regions show values within the 16th and 84th percentiles, i.e., the $\pm 1\sigma$ scatter around the median. From the top-left to the bottom-right, the different panels show the mass-weighted median gas overdensity, temperature, maximum past temperature, pressure, entropy, metallicity, radial peculiar velocity, the mean accretion rate, and the mean mass fraction of hot-mode gas, respectively. The radial profiles at $z=0$ follow the same trends as at $z=2$ (compare Fig.~\ref{fig:haloradz2}), although the pressure, the infall velocity and the accretion rates are lower, the entropy is higher, and the hot mode accounts for a greater fraction of the mass.}
\end{figure*}

Fig.~\ref{fig:haloradz0} is identical to Fig.~\ref{fig:haloradz2} except that it shows profiles for $z=0$ rather than $z=2$. For comparison, the dotted curves in Fig.~\ref{fig:haloradz0} show the corresponding $z=2$ results. As we are again focusing on $10^{11.5}$~M$_\odot<M_\mathrm{halo}<10^{12.5}$~M$_\odot$, the results are directly relevant for the Milky Way galaxy. 

Comparing Fig.~\ref{fig:haloradz2} to Fig.~\ref{fig:haloradz0} (or comparing solid and dotted curves in Fig.~\ref{fig:haloradz0}), we see that the picture for $z=0$ looks much the same as it did for $z=2$. There are, however, a few notable differences. The overdensity profiles hardly evolve, although the difference between the cold and hot modes is slightly smaller at lower redshift. However, a constant overdensity implies a strongly evolving proper density ($\rho \propto (1+z)^3$) and thus also a strongly evolving cooling rate. The large decrease in the proper density caused by the expansion of the Universe also results in a large drop in the pressure and a large increase in the entropy. 

The lower cooling rate shifts the peak of the cold-mode temperature profile from about 2$R_{\rm vir}$ to about $R_{\rm vir}$. While there is only a small drop in the temperature of the hot-mode gas, consistent with the mild evolution of the virial temperature of a halo of fixed mass ($T_{\rm vir} \propto (1+z)$; eq.~[\ref{eqn:virialtemperature}]), the evolution in the median temperature for all gas is much stronger than for the individual accretion modes. While at $z=2$ the overall median temperature only tracks the median hot-mode temperature around $0.5R_{\rm vir} < R< 2 R_{\rm vir}$, at $z=0$ the two profiles are similar at all radii. 

The metallicity profiles do not evolve much, except for a strong increase in the median metallicity of cold-mode gas at $R \gg R_{\rm vir}$. Both the weak evolution of the metallicity of dense gas and the stronger evolution of the metallicity of the cold, low-density intergalactic gas far away from galaxies are consistent with the findings of \citet{Wiersma2011}, who found these trends to be robust to changes in the subgrid physics. 

The absolute, net radial velocities of both the hot- and cold-mode components are smaller at lower redshift, as expected from the scaling of the characteristic velocity ($v_{\rm esc}\propto (1+z)^{1/2}$; eq.~[\ref{eqn:esc}]). 

While at $z=2$ the net accretion rate was higher for the cold mode at all radii, at $z=0$ the hot mode dominates beyond $3R_\mathrm{vir}$. At low redshift the rates are about an order of magnitude lower than at $z=2$. For $R<0.4 R_{\rm vir}$ the net accretion rate is of order 1~M$_\odot$\,yr$^{-1}$, which is dominated by the cold mode, even though most of the mass is in the hot mode. Since a substantial fraction of both the cold- and hot-mode gas inside this radius is outflowing, the actual accretion rates will be a bit higher.

At $z=0$ the fraction of the mass that has been hotter than $10^{5.5}~$K exceeds 50 per cent at all radii and the profile show a broad peak of around 80 per cent at $0.3 < R < R_{\rm vir}$. Thus, hot-mode gas is more important at $z=0$ than at $z=2$, where cold-mode gas accounts for most of the mass for $R \la 0.3 R_{\rm vir}$ and $R\ga 2 R_{\rm vir}$. This is consistent with \citet{Voort2011a}, who investigated the evolution of the hot fraction in more detail using the same simulations.

\section{Conclusions and discussion} \label{sec:concl}

We have used cosmological hydrodynamical simulations from the OWLS project to investigate the physical properties of gas in and around haloes. We paid particular attention to the differences in the properties of gas accreted in the cold and hot modes, where we classified gas that has remained colder (has been hotter) than $10^{5.5}$~K while it was extragalactic as cold-mode (hot-mode) gas.
Note that our definition allows hot-mode gas to be cold, but that cold-mode gas cannot be hotter than $10^{5.5}$~K. We focused on haloes of $10^{12}~$M$_\odot$ at $z=2$ drawn from the OWLS reference model, which includes radiative cooling (also from heavy elements), star formation, and galactic winds driven by SNe. However, we also investigated how the properties of gas near the virial radius change with halo mass, we compared $z=2$ to $z=0$, we measured properties separately for inflowing and outflowing gas, and we studied the effects of metal-line cooling and feedback from star formation and AGN. We focused on mass-weighted median gas properties, but noted that volume-weighted properties are similar to the mass-weighted properties of the hot-mode gas, because most of the volume is filled by dilute, hot gas, at least for haloes with $T_{\rm vir} \ga 10^{5.5}$~K (see Figs.~\ref{fig:dens} and \ref{fig:halo}).

Let us first consider the properties of gas just inside the virial radius of haloes drawn from the reference model at $z=2$ (see Fig.~\ref{fig:halomassz2}). The fraction of the gas accreted in the hot mode increases from 10 per cent for halo masses $M_\mathrm{halo}\sim 10^{10}~$M$_\odot$ to 90 per cent for $M_\mathrm{halo}\sim 10^{13}$~M$_\odot$. Hence, $10^{12}~$M$_\odot$ is a particularly interesting mass scale, as it marks the transition between systems dominated by cold and hot mode gas.

Although the cold streams are in local pressure equilibrium with the surrounding hot gas, cold-mode gas is physically distinct from gas accreted in the hot mode. It is colder ($T < 10^5~$K vs.\ $T\ga T_{\rm vir}$) and denser, particularly for high-mass haloes. While hot-mode gas at the virial radius has a density $\sim 10^2\left <\rho\right >$ for all halo masses, the median density of cold-mode gas increases steeply with the halo mass.  

While the radial peculiar velocity of cold-mode gas is negative (indicating infall) and scales with halo mass like the escape velocity, the median hot-mode velocities are positive (i.e.\ outflowing) for $M_\mathrm{halo} \la 10^{11.5}$~M$_\odot$ and for larger masses they are much less negative than for cold-mode gas. Except for $M_\mathrm{halo} \sim 10^{10.5}$~M$_\odot$, the net accretion rate is positive. For $10^{12} < M_\mathrm{halo} < 10^{13}$~M$_\odot$ the cold- and hot-mode accretion rates are comparable.

While hot-mode gas has a metallicity $\sim 10^{-1}~Z_\odot$, the metallicity of cold-mode gas is typically significantly smaller and displays a much larger spread. The scatter in the local metallicity of cold-mode gas is large because the cold filaments contain low-mass galaxies that have enriched some of the surrounding cold gas. We emphasized that we may have overestimated the median metallicity of cold-mode gas because we find it to decrease with increasing resolution. It is therefore quite possible that most of the cold-mode
gas in the outer halo still has a primordial composition. 

The radial profiles of the gas properties of haloes with $10^{11.5} < M_\mathrm{halo} < 10^{12.5}$~M$_\odot$ revealed that the differences between gas accreted in the cold and hot modes vanishes around $0.1 R_{\rm vir}$ (see Fig.~\ref{fig:haloradz2}), although the radius at which this happens decreases slightly with the resolution. The convergence of the properties of the two modes at small radii is due to catastrophic cooling of the hot gas at $R \la 0.2 R_{\rm vir}$. Interestingly, in the absence of metal-line cooling the hot-mode gas remains hot down to much smaller radii, which suggests that it is very important to model the small-scale chemical enrichment of the circumgalactic gas. While stronger winds do move large amounts of hot-mode gas beyond the virial radius, even AGN feedback is unable to prevent the dramatic drop in the temperature of hot-mode gas inside $0.2R_{\rm vir}$, at least at $z=2$. 

While the density and pressure decrease steeply with radius, the mass-weighted median temperature peaks around $0.5-1.0R_{\rm vir}$. This peak is, however, not due to a change in the temperature of either the cold- or hot-mode gas, but due to the radial dependence of the mass fraction of gas accreted in the hot mode. The hot-mode fraction increases towards the halo, then peaks around $0.5-1.0R_\mathrm{vir}$ and decreases moving further towards the centre. Even outside the halo there
is a significant amount of hot-mode gas (e.g.\ $\sim 30$ per cent at $10R_{\rm vir}$). 

Beyond the cooling radius, i.e.\ the radius where the cooling time equals the Hubble time, the temperature of the hot-mode gas decreases slowly outwards, but the temperature of cold-mode gas only peaks around $2R_{\rm vir}$ at values just below $10^5~$K. The density for which this peak temperature is reached depends on the interplay between cooling (both adiabatic and radiative) and photo-heating. The metallicity decreases with radius and does so more strongly for
cold-mode gas. Near $0.1 R_{\rm vir}$ the scatter in the
metallicity of cold-mode gas is much reduced, although this could be
partly a resolution effect.

The median radial peculiar velocity of cold-mode gas is most negative  around $0.5-1.0R_{\rm vir}$. For the hot mode, on the other hand, it is close to zero around that same radius. Hence, the infall velocity of the cold streams peaks where the hot-mode gas is nearly static, or outflowing if the feedback is very efficient. We note, however, that the scatter in the peculiar velocities is large. For $R\la R_{\rm vir}$ much of the hot-mode gas is outflowing and the same is true for cold-mode gas at $R\sim 0.1R_{\rm vir}$. 

Inside the halo the cold-mode accretion rate increases only slightly with radius ($d\ln \dot{M}/d\ln R \approx 0.4$), indicating that most of the mass is transported to the central galaxy. For the hot-mode, on the other hand, the accretion rate only flattens at $R\lesssim
0.4R_\mathrm{vir}$. This implies that the hot accretion mode mostly feeds the hot halo. However, hot-mode gas that reaches radii $\sim 0.1R_{\rm vir}$ is efficiently transported to the centre as a result of the strong cooling flow. Nevertheless, cold-mode accretion dominates the accretion rate at all radii.

Dividing the gas into inflowing and outflowing components yields results that are very similar to classifying the gas on the basis of its maximum past temperature. This is because inflowing gas is mostly cold-mode and outflowing gas is mostly hot-mode. The situation is, however, different for high-mass haloes ($M_\mathrm{halo}\ga 10^{12.5}$~M$_\odot$). Because the two accretion modes bring similar amounts of mass into these haloes, the properties of the infalling gas are intermediate between those of the cold and hot accretion modes.

When expressed in units of the mean density of the universe, the $z=0$ density profiles (with radius expressed in units of the virial radius) are very similar to the ones at $z=2$. The same is true for the metallicities and temperatures, although the peak temperatures shift to slightly smaller radii (again normalized to the virial radius) with time. A fixed overdensity does imply that the proper density evolves as $\rho\propto (1+z)^3$, so the pressure (entropy) are much lower (higher) at low redshift. Infall velocities and accretion rates are also significantly lower, while the fraction of gas accreted in the hot mode is higher. The difference in the behaviour of the two accretion modes is, however, very similar at $z=0$ and $z=2$.  

Although there are some important differences, the overall properties of the two gas modes are very similar between our different simulations, and are therefore insensitive to the inclusion of metal-line cooling and galactic winds. Without SN feedback, the already dense cold-mode gas reaches even higher densities, the metallicity of hot-mode gas in the outer halo is much lower. Without metal-line cooling, the temperature at radii smaller than $0.2R_{\rm vir}$ is much higher. With strong
SN feedback or AGN feedback, a larger fraction of the
gas is outflowing, the outflows are faster, and the peak in the fraction of the gas that is accreted in the hot mode peaks at a much larger radius (about $3R_{\rm vir}$ instead of $0.5-1.0 R_{\rm vir}$).

\citet{Keres2011} have recently shown that some of the hot gas properties depend on the numerical technique used to solve the hydrodynamics. They did, however, not include metal-line cooling and feedback. They find that the temperatures of hot gas (i.e.\ $T>10^5$~K) around $10^{12-13}$~M$_\odot$ haloes are the same between the two methods, but that the median density and entropy of $10^{12}$~M$_\odot$ haloes is somewhat different, by less than a factor of two. These differences are comparable to the ones we find when using different feedback models. They also show that the radial velocities are different at small radii. Our results show that including metal-line cooling decreases the radial velocities, whereas including supernova or AGN feedback increases the radial velocities significantly. Even though there are some differences, the main conclusions of this work are unchanged. The uncertainties associated with the subgrid implementation of feedback and the numerical method are therefore unlikely to be important for our main conclusions.

Cold (i.e.\ $T \ll 10^5$~K) outflows are routinely detected in the form of blueshifted interstellar absorption lines in the rest-frame UV spectra of star-forming galaxies \citep[e.g.][]{Weiner2009, Steidel2010, Rubin2010, Rakic2011a}. This is not in conflict with our results, because we found that the outflowing gas spans a very wide range of temperatures and because the detectable UV absorption lines are biased towards colder, denser gas. Additionally, the results we showed are mass-weighted, but if we had shown volume-weighted quantities, the outflow fractions would have been larger. The inflowing material has smaller cross-sections and is therefore less likely to be detected \citep[e.g.][]{FaucherKeres2011, Stewart2011a}. 

How can we identify cold-mode accretion observationally? The two modes only differ clearly in haloes with $T_{\rm vir} \gg 10^5$~K ($M_{\rm halo} \gg 10^{10.5} ~{\rm M}_\odot ((1+z)/3)^{-3/2}$), because photo-ionization by the UV background radiation ensures that all accreted gas is heated to temperatures up to $\sim 10^5$~K near the virial radius.  Near the central galaxy, $R\la 0.1R_{\rm vir}$, it is also difficult to distinguish the two modes, because gas accreted in the hot mode is able to cool. 

In the outer parts of sufficiently massive haloes the properties of the gas accreted in the two modes do differ strongly. The cold-mode gas is confined to clumpy filaments that are approximately in pressure equilibrium with the diffuse, hot-mode gas. Besides being colder and denser, cold-mode gas typically has a much lower metallicity and is much more likely to be infalling. However, the spread in the properties of the gas is large, even for a given mode and a fixed radius and halo mass, which makes it impossible to make strong statements about individual gas clouds. Nevertheless, it is clear that most of the dense ($\rho \gg 10^2 \left <\rho\right >$) gas in high-mass haloes ($T_{\rm vir}\ga 10^6$~K) is infalling, has a very low metallicity and was accreted in the cold-mode. 

Cold-mode gas could be observed in UV line emission if we are able to detect it in the outer halo of massive galaxies. Diffuse Lyman-$\alpha$ emission has already been detected \citep[e.g.][]{Steidel2000, Matsuda2004}, but its interpretation is complicated by radiative transfer effects and the detected emission is more likely scattered light from central H\,\textsc{ii} regions (e.g.\ \citealt{Furlanetto2005, Faucher2010, Steidel2011, Hayes2011, Rauch2011}, but see also \citealt{Dijkstra2009, Rosdahl2012}). Metal-line emission from ions such as C~\textsc{iii}, C~\textsc{iv}, Si~\textsc{iii}, and Si~\textsc{iv} could potentially reveal cold streams, but current facilities do not probe down to the expected surface brightnesses \citep{Bertone2010b,Bertone2012}. 

The typical temperatures ($T\sim 10^4$~K) and densities ($\rho \ga 10^2 \left <\rho\right >$) correspond to those of strong quasar absorption lines systems. For example, at $z=2$ the typical H\,\textsc{i} column density is\footnote{For $N_{\rm H\,\textsc{i}} \ga 10^{18}~{\rm cm}^{-2}$ the relation is modified by self-shielding, see e.g.\ \citealt{Schaye2001b,Altay2011}.} $N_{\rm H\,\textsc{i}} \sim 10^{16} ~{\rm cm}^{-2}(\rho/[10^2 \left <\rho\right >])^{3/2}$ \citep{Schaye2001a} with higher column density gas more likely to have been accreted in the cold mode. At low redshift the H\,\textsc{i} column densities corresponding to a fixed overdensity are about 1--2 orders of magnitude lower \citep{Schaye2001a}.

Indeed, simulations show that Lyman limit systems (i.e.\ $N_{\rm H\,\textsc{i}} > 10^{17.2} ~{\rm cm}^{-2}$) may be used to trace cold flows \citep{FaucherKeres2011,Fumagalli2011a,Voort2011c} and \citet{Voort2011c} have demonstrated that cold-mode accretion is required to match the observed rate of incidence of strong absorbers at $z=3$. Many strong QSO absorbers also tend to have low metallicities \citep[e.g.][]{Ribaudo2011,Giavalisco2011,Fumagalli2011b}, although it should be noted that metallicity measurements along one dimension may underestimate the mean metallicities of three-dimensional gas clouds due to the expected poor small-scale metal mixing \citep{Schaye2007}. We also note that most Lyman limit systems are predicted to arise in or around haloes with masses that are much lower than required for the presence of stable accretion shocks near the virial radius, so that they will generally not correspond to cold streams penetrating hot, hydrostatic haloes \citep{Voort2011c}. To study those, it is therefore more efficient to target sight lines to QSOs close to massive foreground galaxies \cite[e.g.][]{Rakic2011b,FaucherKeres2011,Stewart2011a,Kimm2011}.

\section*{Acknowledgements}

We would like to thank the referee, Daniel Ceverino, for helpful comments and Robert Crain, Alireza Rahmati and all the members of the OWLS team for valuable discussions. The simulations presented here were run on Stella, the LOFAR BlueGene/L system in Groningen, on the Cosmology Machine at the Institute for Computational Cosmology in Durham as part of the Virgo Consortium research programme, and on Darwin in Cambridge. This work was sponsored by the National Computing Facilities Foundation (NCF) for the use of supercomputer facilities, with financial support from the Netherlands Organization for Scientific Research (NWO), also through a VIDI grant, and from the Marie Curie Initial Training Network CosmoComp (PITN-GA-2009-238356).

\bibliographystyle{mn2e}
\bibliography{propertiesRvir}

\bsp

\appendix
\section{Resolution tests}

\begin{figure*}
\center
\includegraphics[scale=0.62]{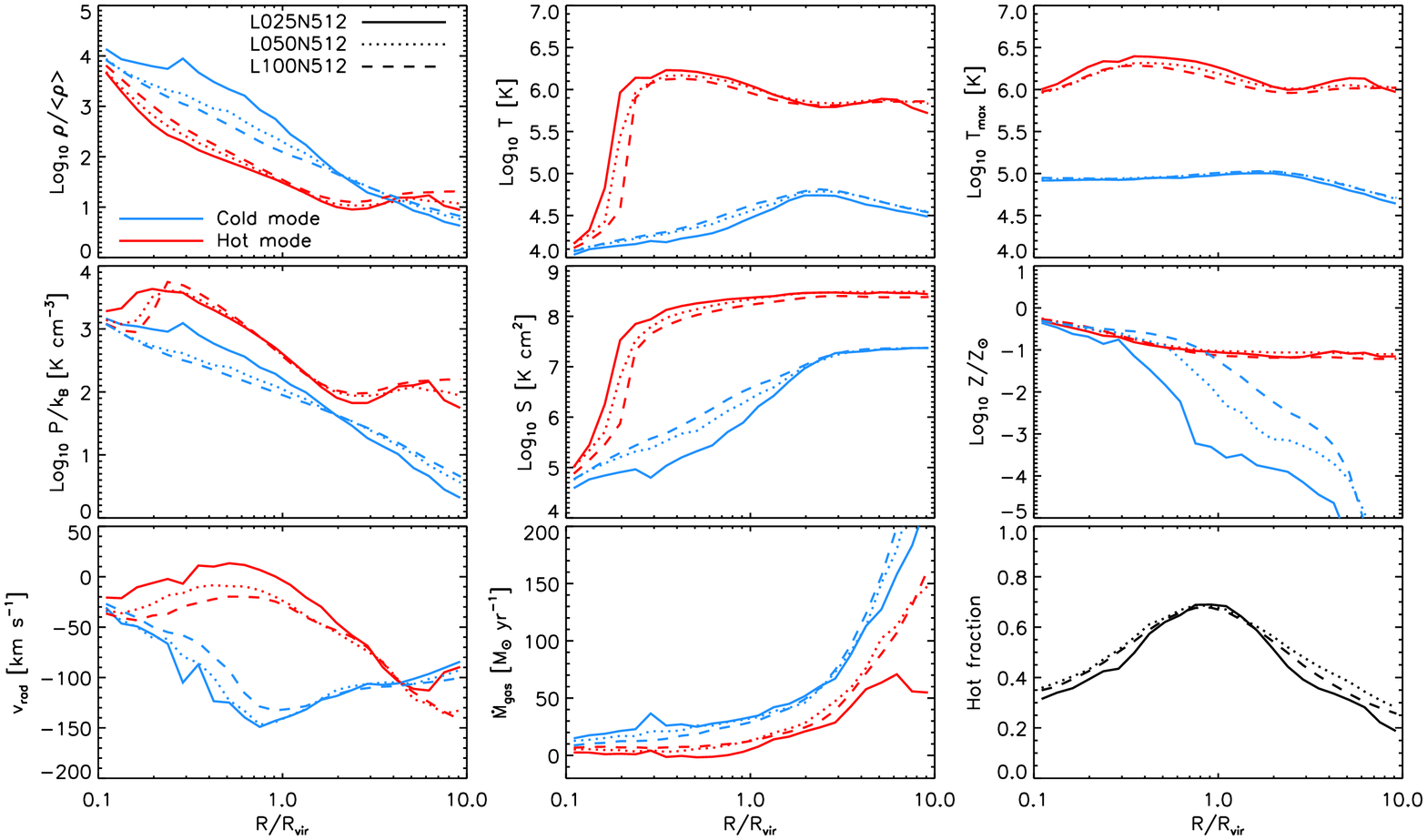}
\caption {\label{fig:haloradres} Convergence of gas profiles with resolution. All profiles are for haloes with $10^{11.5}$~M$_\odot<M_\mathrm{halo}<10^{12.5}$~M$_\odot$ at $z=2$. Results are shown for simulations \emph{REF\_L025N512} (solid), \emph{REF\_L050N512} (dotted), and \emph{REF\_L050N512} (dashed) and for both the hot (red) and cold (blue) modes. The three simulations vary by a factor of 64 (8) in mass (spatial) resolution. From the top-left to the bottom-right, the different panels show the mass-weighted median gas overdensity, temperature, maximum past temperature, pressure, entropy, metallicity, radial peculiar velocity, the mean accretion rate, and the mean mass fraction of hot-mode gas, respectively.}
\end{figure*}
\begin{figure*}
\center
\includegraphics[scale=0.62]{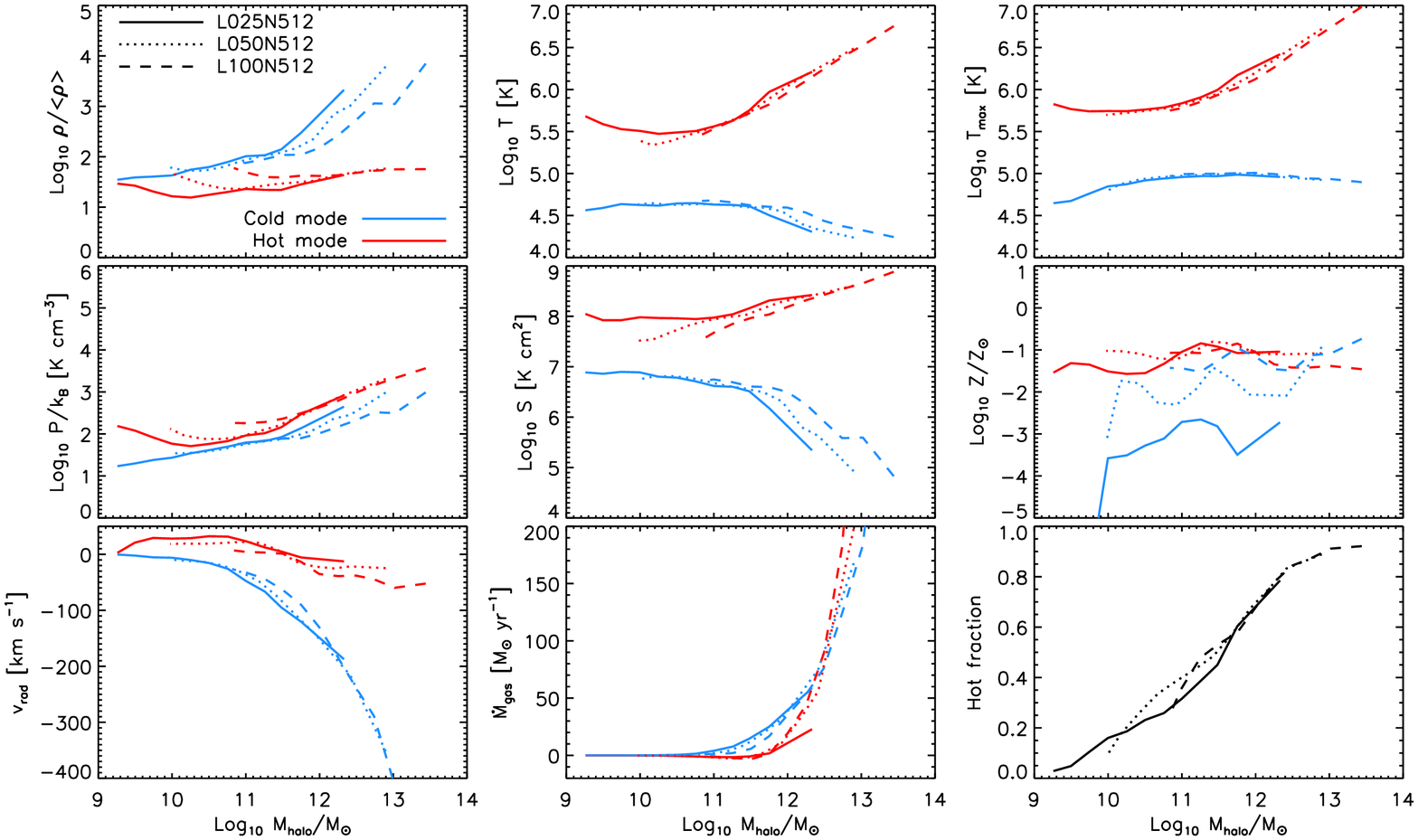}
\caption {\label{fig:halomassres} Convergence of the properties of gas at $0.8 R_\mathrm{vir}<R<1 R_\mathrm{vir}$ as a function of halo mass with resolution. Panels and curves as in Fig.~\ref{fig:haloradres}.}
\end{figure*}

We have checked (but do not show) that the results presented in this work are converged with respect to the size of the simulation volume if we keep the resolution fixed. The only exception is hot-mode gas at $R>5R_\mathrm{vir}$, for which the radial velocities and accretion rate require a box of at least 50$h^{-1}$Mpc on a side (which implies that our fiducial simulation is sufficiently large). 

Convergence with resolution is, however, more difficult to achieve. In Figs.~\ref{fig:haloradres} and~\ref{fig:halomassres} we show again the radial profiles and mass dependence of the halo properties for the hot- and cold-mode components at $z=2$. Shown are three different simulations of the reference model, which vary by a factor of 64 (8) in mass (spatial) resolution. All trends with radius and halo mass are very similar in all runs, proving that most of our conclusions are robust to changes in the resolution. Below we will discuss the convergence of hot- and cold-mode gas separately and in more detail.

The convergence is generally excellent for hot mode gas. As the resolution is increased, the density of hot-mode gas decreases slightly and the temperature drop close to the halo centre shifts to slightly smaller radii, which also affects the pressure and entropy. There is a small upturn in the density of hot-mode gas at the virial radius as we approach the halo mass corresponding to the imposed minimum of 100 dark matter particles, showing that we may have to choose a minimum halo mass that is a factor of 5 higher for complete convergence. The radial peculiar velocity increases slightly with resolution, causing the hot-mode accretion rate to decrease. 

Convergence is more difficult to achieve for cold-mode gas. The density of cold-mode gas inside haloes, and thereby also the difference between the two modes, increases with the resolution. The pressure of the cold-mode gas also increases somewhat with resolution, which leads to a smaller difference with the pressure of hot-mode gas. 
The cold-mode radial velocity becomes more negative, increasing the difference between the two modes. 

The median metallicity of the cold-mode gas decreases strongly with increasing resolution for $R>0.3R_\mathrm{vir}$. The metallicity difference between the two modes therefore increases, although the distributions still overlap (not shown), and the radius at which the metallicities of the two modes converge decreases. In fact, the convergence of the median metallicity of cold-mode gas is so poor that we cannot rule out that it would tend to zero at all radii if we keep increasing the resolution. 

If we use particle metallicities rather than SPH smoothed metallicities (see Section~\ref{sec:metal}), then the median metallicity of both modes is lower and the median cold-mode metallicity plummets to zero around the virial radius (not shown). The decrease in metallicity with increasing resolution is in that case less strong, but still present. The unsmoothed hot-mode metallicities are also not converged and lower than the (converged) smoothed hot-mode metallicities, but they increase with increasing resolution. The difference between the two modes therefore increases with resolution. 

With increasing resolution, the radial velocities of cold-mode gas become slightly more negative within the halo. The net accretion rates and hot fractions are converged. 

Even though some properties are slightly resolution dependent, or strongly so for the case of the metallicity of cold-mode gas, all our conclusions are robust to increases in the numerical resolution.

\label{lastpage}

\end{document}